\documentclass[]{aastex631}
\shorttitle{Atmospheric Circulation on Sub-Neptunes}
\shortauthors{Landgren et al.}
\graphicspath{{./}{figures/}}

\begin{document}


\title{A Shallow Water Model Exploration of Atmospheric Circulation on Sub-Neptunes: Effects of Radiative Forcing and Rotation Period}

\correspondingauthor{Ekaterina Landgren}
\email{ek672@cornell.edu}

\author[0000-0001-6029-5216]{Ekaterina Landgren} 
\affil{Cooperative Institute for Research in Environmental Sciences\\
University of Colorado Boulder\\
Ekeley Science Building\\
Boulder, CO 80302}

\author[0000-0003-4325-8047]{Alice Nadeau} 
\affil{Department of Mathematics \\
Cornell University\\
310 Malott Hall\\
Ithaca, NY 14853}

\author[0000-0002-8507-1304]{Nikole Lewis} 
\affil{Department of Astronomy and Carl Sagan Institute \\
Cornell University\\
122 Sciences Drive\\
Ithaca, NY 14853}

\author[0000-0003-3759-9080]{Tiffany Kataria} 
\affil{Jet Propulsion Laboratory \\
California Institute of Technology\\
Pasadena, CA 91109}

\author[0000-0001-8993-3808]{Peter Hitchcock} 
\affil{Department of Earth and Atmospheric Sciences \\
Cornell University\\
2134 Snee Hall\\
Ithaca, NY 14853}



\begin{abstract}
Sub-Neptune type exoplanets are abundant in our galaxy yet have no solar system analogs. They exist in a broad range of stellar forcing and rotational regimes that are distinctly different from solar system planets and more commonly studied hot Jupiters. Here we present simulations that explore global atmospheric circulation of sub-Neptunes generated with a two-dimensional shallow-water model, SWAMPE. We explore the circulation regimes of synchronously rotating sub-Neptunes with a focus on the interaction of planetary rotation rate and radiative timescale in a variety of stellar insolations. In highly irradiated, short-timescale regimes, our models exhibit high day-night geopotential contrasts. As the timescales become longer, the geopotential contrasts and longitudinal variability decrease, while temporal variability increases. The transition from day-to-night flow to jet-dominated flow is primarily driven by the radiative timescale. Strong- and medium-forcing regimes exhibit transitions between day-to-night flow and jet-dominated flow at similar points in the parameter space. Weak-forcing regime differs due to comparatively stronger rotational effects. Planetary rotation period dominates in determining equator-to-pole geopotential contrast. Our simulations exhibit higher time variability when either radiative timescale or rotation period is long.
\end{abstract}



\section{Introduction} \label{sec:intro}

Despite being some of the most common planet types in our galaxy \citep{handbook2018}, sub-Neptunes, planets 1.6-4 times the radius of Earth, have no analogs in the solar system. This population of planets, which can be broken into two sub-groups, rocky super-Earths and gaseous mini-Neptunes, are prime candidates for atmospheric modeling that can reveal a variety of physical phenomena. In order to answer questions about potential habitability of terrestrial exoplanets, it is important to develop a robust understanding of the dynamical processes that can take place in the atmospheres of these small planets. Sub-Neptunes exist in a wide phase space in terms of temperature (most of them falling in the 400-1200~K equilibrium temperature range \citep{crossfield2017trends}), and are predicted to have a broad range of atmospheric chemical compositions \citep{kite2020}. Investigating the atmospheric circulation and the mechanisms underlying global scale flows of planets across the entire sub-Neptune parameter space and providing observational predictions can be time-consuming with high-complexity three-dimensional (3D) general circulation models, especially those that try to treat the full range of chemical and physical processes taking place in their atmospheres. On the other end of model complexity, many one-dimensional (1D) models can be insufficient in capturing the inherently 3D processes taking place in their atmospheres \citep[e.g.][]{feng2016impact,kataria2016,line2016}.
 
 For these reasons, two-dimensional (2D) shallow-water models are useful tools that can capture key dynamical processes despite being more simplified than 3D models. Such models represent the dynamics of a thin layer of a fluid of constant density and variable thickness. The shallow-water equations are a coupled system of horizontal momentum and mass conservation, which govern the evolution of the zonal and meridional velocity components as well as the layer thickness. These equations are typically solved numerically as a function of longitude, latitude, and time. For a description of the shallow-water model focused on application to exoplanetary atmospheric circulation, see, e.g. \citet{showman2010atmospheric}.

There is a rich heritage of shallow-water models that have been used to study solar system planets.
 For example, shallow-water 2D GCMs have been used to study topography variation on Earth \citep{ferrari2004new}, \citep{kraucunas2007tropical}, planetary-scale waves on Venus \citep{iga2005shear},  latitudinal transport and superrotation on Titan \citep{luz2003latitudinal}, and polar vortex dynamics on Jupiter, Saturn, Uranus, and Neptune \citep{brueshaber2019dynamical} as well as on Mars \citep{rostami2017understanding}. Shallow-water models have been used to model the atmospheric phenomena of objects beyond the solar system as well. 
 \citet{perez2013atmospheric} explore the interaction of radiative and drag timescales using a 2D model,  \citet{zhang2014} explore the circulation of brown dwarfs,  \citet{penn2018} apply the shallow-water framework to terrestrial exoplanets, \citet{hammond2018} use it to consider tidally locked exoplanets, and \citet{ohno2019} model nonsynchronized, eccentric-tilted exoplanets.

The atmospheric circulation of sub-Neptunes have been probed in several studies using mainly 3D GCMs. \citet{zhang2017effects} use a 3D GCM to explore the effects of bulk composition on the atmospheres of sub-Neptunes, and find that the effect of molecular weight dominates in determining day-night temperature contrast and the width of the superrotating jet. \citet{Wang2020, christie2022} study the atmospheric circulation of the sub-Neptune GJ 1214 b using LMDZ GCM and Met Office’s UM GCM, respectively, and find that a long convergence time is needed to simulate long radiative timescale, and that high metallicity and clouds with large vertical extents are needed to match synthetic observations. \citet{innes2022} investigate tidally locked, temperate (with Earth-like insolation) sub-Neptunes using the ExoFMS GCM and find that these atmospheres exhibit weak horizontal temperature gradients, high-latitude cyclostrophic jets, and equatorial superrotation at low pressures. However, these previous studies of sub-Neptunes are limited in their scope, either to singular planets or to narrow phase spaces. Using 2D models to explore the circulation of sub-Neptunes would enable a wider exploration of this parameter space.

 In this study, we use an open-source 2D shallow water model, \texttt{SWAMPE} \citep{landgren2022swampe}, to investigate the atmospheric dynamics of sub-Neptunes. We contribute to the growing literature on sub-Neptunes by systematically investigating the interaction of planetary rotation period and radiative forcing and their effect on planetary atmospheric dynamics. Our goal is to explore the atmospheric circulation of sub-Neptunes and to identify the key differences in the dominant atmospheric circulation patterns of this type of planets compared to hotter tidally-locked Jovian atmospheres. While idealized, the shallow-water framework is well-suited for the analysis of the dynamics of large-scale atmospheric flow, which is the focus of this work. 

The paper is organized as follows. In Section \ref{sec:model}, we introduce our dynamical model and formulate radiative forcing. In Section \ref{sec:validation}, we validate our model implementation against a hot Jupiter regime based on the planetary parameters for HD 189733b. In Section \ref{sec:params}, we motivate the phase space we choose to explore in this investigation of sub-Neptunes. In Section \ref{sec:results}, we present our simulation results and examine the global atmospheric flows and horizontal winds for strongly forced, medium forced, and weakly forced sub-Neptunes. In Section \ref{sec:discussion}, we discuss our simulation results, highlighting general trends in heating patterns and zonal winds as well as phenomena of interest such as idiosyncratic spin-up behavior, oscillations, and temporal variability. Finally, we conclude in Section \ref{sec:conclusion}.

\section{Methods}
\label{sec:methods}
\subsection{Model} \label{sec:model}

  We study atmospheric circulation on sub-Neptunes using an idealized shallow-water model, \texttt{SWAMPE} (Shallow-Water Atmospheric Model in Python for Exoplanets, \citet{landgren2022swampe}). The atmosphere is assumed to be a fluid that is incompressible and hydrostatically balanced. For a derivation of the shallow water equations from the continuity equation and the equation of motion, see, e.g. \citet{kaper2013mathematics}. As the name suggests, the shallow-water simplification of the Navier-Stokes equations assumes that the fluid (in our case, the planetary atmosphere) is spread in a thin layer (i.e. the horizontal scale is much larger than the vertical scale). The simplification of the full dynamics comes with several limitations. The model omits baroclinic dynamics, which affect heat redistribution, especially for fast rotators \citep{penn2017}. The shallow-water system employs a simple forcing term to account for stellar insolation and lacks the complexity of a more realistic radiative transfer scheme: the details of absorption, emission, and scattering are omitted. Additionally, the model does not incorporate frictional diffusivity, thermal radiation to space, or stratification.   This idealized system can nevertheless capture large-scale phenomena on a sphere in response to gravitational and rotational accelerations.  Within this framework, the wind patterns exhibited by the model can be explained by considering the interaction of stationary, planetary-scale waves with the mean flow \citep[e.g.][]{showman2011equatorial, showman2012doppler}. These interactions are robust and the dynamics captured by the shallow-water model are likely to translate into the full dynamics of the atmosphere. The Python code developed for this work is based on the method described by \citet{hack1992description} and is available for download on Github.\footnote{ \url{https://github.com/kathlandgren/SWAMPE}}

  The idealized shallow-water system models the atmosphere as two layers: a buoyant upper layer at constant density $\rho_{\rm upper}$ and an infinitely deep convective layer at a higher constant density $\rho_{\rm lower}$. Sub-Neptunes are likely to have deep atmospheres ($>10^3$ km atmospheres, see e.g. \citet{kite2020}), and so the shallow-water framework is appropriate for modeling these types of exoplanets. Vertical transport is simplified in the model, represented only as mass transfer between layers. Hence we focus on modeling horizontal velocities in the upper layer.

The equations governing the top layer are: 
    
    \begin{align}
    \label{eq-hor-momentum}
        \frac{d\mathbf{V}}{dt}+f\mathbf{k}\times\mathbf V+\nabla\Phi&=F_{\mathbf{V}},\\
    \label{eq-mass-cont}
        \frac{d\Phi}{dt}+\Phi\nabla\cdot\mathbf{V}&=F_{\Phi},
    \end{align}

where $\mathbf V=u\mathbf{i}+v\mathbf{j}$ is the horizontal velocity vector and $\mathbf i$ and $\mathbf j$ are the unit eastward and northward directions, respectively. The quantities $F_{\mathbf{V}}$ and $F_{\Phi}$ refer to the forcing terms in the momentum equation and mass continuity equation, respectively. In this work, these terms are tied to stellar insolation, but in general they represent any source or sink terms. The free surface geopotential is given by $\Phi \equiv gh$ where $g$ is the gravitational acceleration and $h$ is geopotential height. The geopotential height represents the height of the pressure surface of the modeled upper layer. The geopotential can be separated into the constant component $\overline{\Phi}=gH$ (where $H$ is the scale height) and the time-dependent deviation $\Phi$. We follow this convention here \citep[following, e.g.][]{hack1992description, perez2013atmospheric}.

The geopotential forcing is represented by linear relaxation to the local radiative equilibrium state $\Phi_{\text{eq}}$, given by:
 \begin{equation}
     \Phi_{\text{eq}}= \begin{cases} \overline{\Phi}+\Delta \Phi_{\text{eq}} \cos \lambda \cos \phi & \text{on the dayside},\\
     \overline{\Phi} & \text{on the nightside,}
   \end{cases}
   \label{eq:Q}
 \end{equation}
 where $\Delta \Phi_{\text{eq}}$ is the maximal day-night contrast assuming the thickness of the upper layer in radiative equilibrium. The ratio $\Delta \Phi_{\text{eq}}/\overline{\Phi}$ serves as a proxy for the equilibrium day-night temperature contrast. The radiative equilibrium thickness is highest at the substellar point, which occurs at latitude $0^{\circ}$, longitude $0^{\circ}$. This equation represents the radiative heat the planet receives from its star, hence the forcing is only present on the dayside. Here we assume that sub-Neptune planets are orbiting close to their host stars and synchronously rotating with the rotational period equal to the orbital period (1:1). The bulk of the known sub-Neptune population has orbital periods of less than 30 days \citep{bean2021nature} that place them generally within the tidal locking distance of their host stars \citep[e.g][]{barnes2017tidal, leconte2015}.  Hence the substellar point and the radiative equilibrium as a function of latitude and longitude are constant in time in our model.
 Linear relaxation to the local radiative equilibrium takes the following form:
 \begin{equation}
     F_{\Phi}=\dfrac{\Phi_{\text{eq}}-\Phi}{\tau_{\text{rad}}}\equiv Q,
 \end{equation}
    where $\tau_{\text{rad}}$ is the radiative time scale.  In this study, the timescale $\tau_{\rm rad}$ is a free parameter. We give an overview of the values of $\tau_{\rm rad}$ we sample in Section \ref{sec:params}. If the geopotential is below the local radiative equilibrium, the forcing will act to increase it, and if the geopotential is above the local radiative equilibrium, the forcing will act to decrease it. In the shallow-water framework, regions of higher geopotential correspond to regions of heating, since mass is transferred into the upper layer in the presence of radiative forcing.
   
 In the momentum equation, the mass transfer between the two layers in the model is governed by the following expression:
   
   \begin{equation}
   F_{\mathbf{V}}=\begin{cases} 
      -\dfrac{Q\mathbf{V}}{\Phi} & Q> 0, \\
      0 & Q\leq 0 .
   \end{cases}
\end{equation}
This term captures the effect of momentum advection from the lower layer on the upper layer.  The term $F_{\mathbf{V}}$ captures the mass transferred from the lower layer to the upper layer through heating ($Q>0$). The mass transfer from the upper layer to the lower layer due to cooling, on the contrary, does not affect the specific momentum of the upper layer, and so $F_{\mathbf{V}}$ is zero wherever $Q<0$. The expression for $F_{\mathbf{V}}$ used here corresponds to the term $\mathbf{R}$ used in \citet{perez2013atmospheric,showman2011equatorial,showman2012,shell2004abrupt}.

 We solve Equations \ref{eq-hor-momentum} and \ref{eq-mass-cont} using a method based on \citet{hack1992description} implemented in Python. The model employs a spectral method with a global, spherical geometry, which performs differentiation using a truncated sequence of spherical harmonics. The non-linear terms are evaluated in physical space, and are then transformed into the wavenumber (spectral) space using a spectral transform, which is a composition of the Fast Fourier Transform and Gaussian quadrature. In wavenumber space, linear terms and derivatives are computed. Time-stepping is performed and prognostic variables are evaluated in spectral space. The prognostic variables are then transformed back into physical space, where the diagnostic and non-linear terms are evaluated. 
 
 Following \citet{hack1992description}, we use Gaussian latitudes ($\mu=\sin \phi$), uniformly spaced longitude, and $U=u \cos \phi$, $V=v \cos \phi$ for zonal and meridional winds, since \citet{robert1966integration} showed that the variables $U$ and $V$ are better suited to computations in spectral space. While the quantities of interest are the geopotential $\Phi$ and the horizontal wind vector field $\mathbf{V}$, we rely on the equivalent vorticity/divergence form of the shallow-water equations, where the prognostic variables are geopotential $\Phi$, absolute vorticity $\eta$, and divergence $\delta$. In this form, the shallow-water equations have terms that have simple representations in spectral space. The zonal and meridional wind components $U$ and $V$ become diagnostic variables, albeit readily determined from $\eta$ and $\delta$ at any time step. In physical space, the variables can be represented as functions of longitude, latitude, and time: prognostic variables $\Phi(\lambda,\phi,t)$, $\eta(\lambda,\phi,t)$, $\delta(\lambda,\phi,t)$, and diagnostic variables $U(\lambda,\phi,t)$, $V(\lambda,\phi,t)$. Note that there is no explicit dependence on pressure, since the shallow-water system is a barotropic formulation. Readers interested in the derivation of the vorticity/divergence form and the diagnostic relationship should consult, e.g. \citet{hack1992description}. 
  
 We use the spectral truncation T42, corresponding to a spatial resolution of $128\times64$ in longitude and latitude, with $2.81^{\circ}$ cell size. Given that we wish to study global-scale flows in sub-Neptune atmospheres, we opt to use the truncation which provides the best balance between computational efficiency and spatial resolution. Previous studies using GCMs and shallow-water models agree that the dynamics on tidally-locked exoplanets are global in scale \citep[e.g.][]{showman2011equatorial,showman2015three-dimensional, heng2011, perez2013atmospheric, zhang2017effects}. We provide scale analysis for the simulations in our ensemble in Section \ref{sec:discussion}. The equations are integrated using a modified Euler's method time-stepping scheme, rather than a more common semi-implicit scheme, due to stability considerations \citep[following][]{langton2008atmospheric}.
 We apply two filters to further ensure numerical stability: the modal splitting filter \citep{hack1992description}, and a $\nabla^6$ hyperviscosity filter \citep[based on][]{gelb2001spectral}. Models are integrated from an initially flat layer (no deviation from $\overline{\Phi}$) and zero winds. We select a time step that meets the CFL condition \citep{boyd2001chebyshev}. In general, simulations of highly irradiated planets require a shorter time step due to higher wind speeds. The time steps used in our simulations range from 30 s to 180 s. Each model is integrated for 1000 Earth days. We monitor the RMS windspeeds of each simulation to ensure that they reach a steady state, which typically occurs between 50 and 100 simulated days, but may take longer for simulations of weakly forced planets.

   Our model is consistent with the shallow-water framework as it has been previously applied to hot Jupiters. Specifically, the momentum advection term $F_{\mathbf{V}}$ matches the term $\mathbf{R}$ used in \citet{shell2004abrupt,showman2011equatorial,showman2012doppler}. The momentum equation \eqref{eq-hor-momentum} and the mass-continuity equation \eqref{eq-mass-cont} are consistent with the form given in \citet{vallis2017atmospheric}. These equations also reduce to the form given in \citet{perez2013atmospheric} for a regime where the atmospheric drag timescale $\tau_{\rm drag}=\infty$, equivalent to the absence of atmospheric drag. \citet{perez2013atmospheric} have explored how the addition of atmospheric drag interacts with the radiative timescale. Since the focus of this work is on the interplay of radiative timescale and rotation period, which has not yet been explored, we have chosen to omit the drag term. We assume that the modeled layer is sufficiently deep that there is no interaction with the planetary surface. Note also that while \citet{perez2013atmospheric} use the geopotential height as the prognostic variable, the form used in our model is equivalent since the geopotential height and the geopotential differ only by a factor of reduced gravity $g$. To demonstrate the agreement of our model with previous implementations, we model a hot Jupiter in the following section.

\subsection{Model validation}
\label{sec:validation}

\begin{figure}
  \centering
  \includegraphics[width=0.8\linewidth]{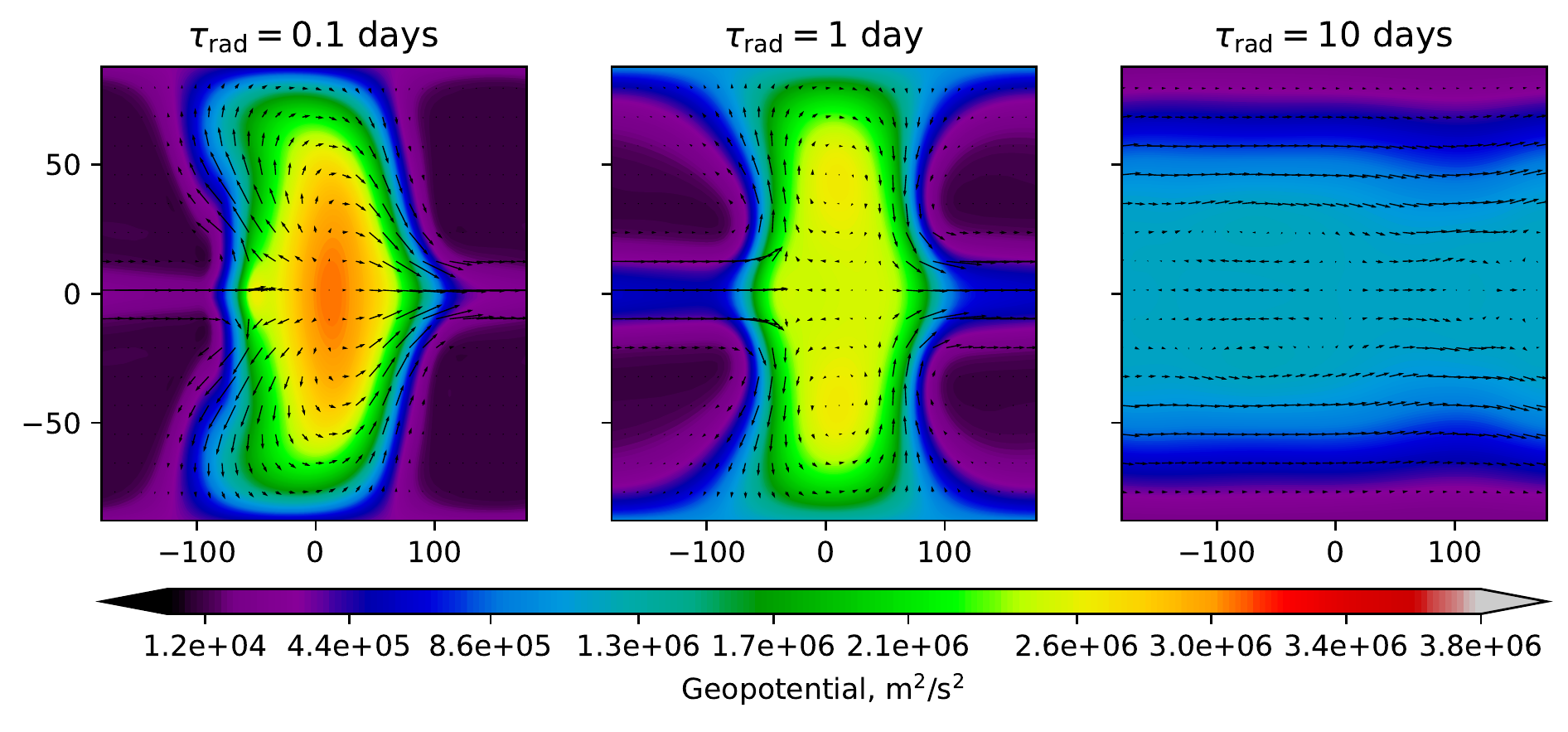}
  \caption{Recreation of a hot Jupiter regime in Figure 3 in \cite{perez2013atmospheric}. Shown in the figure are maps of steady-state geopotential contours and the wind vector fields for the equilibrated solutions of the shallow-water model. The planetary parameters are as follows: planetary radius $a=8.2\times 10^7$ m, planetary rotation rate $\Omega=3.2\times10^{-5}$ rad/s (equivalent to to $P_{\rm rot}=2.27$ days), no drag, reference geopotential $\overline{\Phi}=4 \times 10^6$, for strong radiative forcing ($\Delta \Phi_{\rm eq}/\overline{\Phi}=1$).  Three radiative time scales are depicted: $\tau_{\rm rad=0.1}$ days (left panel), $\tau_{\rm rad=1}$ day (center panel), and $\tau_{\rm rad=10}$ days (right panel). We have subtracted the constant value of $\overline{\Phi}$ from each panel. Panels share the same color scale for geopotential, but winds are normalized independently in each panel. The substellar point is located at $(0^{\circ},0^{\circ})$ for each panel.}
  \label{PBS-recreation}
\end{figure}

    In order to validate our model, we replicate three hot Jupiter regimes from \citet{perez2013atmospheric}, shown in Figure \ref{PBS-recreation}, which matches the three middle panels in the top row of Figure~3 in \citet{perez2013atmospheric}. While the model we are replicating includes the effects of atmospheric drag, in this regime, $\tau_{\rm drag}=\infty$, which corresponds to the absence of drag. We have chosen the color bar to match that of Figure 3 in \citet{perez2013atmospheric}. The other parameters here are $\Delta \Phi_{\rm eq}/\overline{\Phi}=1$ (high contrast regime), geopotential at scale height $\overline{\Phi}=gH=4 \times 10^6$ m$^2$ s$^{-2}$, the planetary radius $a=8.2\times 10^7$ m and the rotation frequency $\Omega=3.2\times10^{-5}$ s$^{-1}$ (equivalent to $P_{\rm rot}=2.27$ days) were selected in \citet{perez2013atmospheric} based on those of HD~189733b. Our simulations match the expected behavior, replicating the transition from shorter $\tau_{\rm rad}$, where the planetary behavior resembles the radiative forcing profile accompanied by a day-to-night flow, to longer $\tau_{\rm rad}$, where the longitudinal gradients are small and the structure exhibits a hot equatorial band.

\subsection{Parameter regimes}
\label{sec:params}

   The modeling phase space of sub-Neptunes is vast. For example, the population of sub-Neptunes detected by \textit{Kepler} exhibits orbital periods ranging from $0.4$ to $300$ Earth days \citep{bean2021nature}. These planets exhibit a variety of possible atmospheric compositions and orbit a range of host stars. Therefore we select to vary the radiative timescale to capture the breadth of possible atmospheric compositions, and vary the intensity of incoming stellar radiation to model the possible range of stellar host types and orbital distances. Under the constraint of synchronous rotation, which necessitates planets to be orbiting within the tidal locking distance, we focus on the interplay of rotation period with the radiative timescale and the strength of insolation.
   
   While sub-Neptune planets range in size from 1.6 to 4 times the radius of Earth, here we select a representative sub-Neptune planetary radius equal to  three Earth radii: $a=3a_{\oplus}=1.91 \times 10^7$ m \citep{bean2021nature}.  The reduction in radius compared to a Jupiter-sized planet $R_J=8.2 \times 10^7$ m leads to significant changes in the atmospheric circulation by affecting the advective timescale and the vorticity gradient. The advective timescale $\tau_{\rm adv}\approx a/U$ decreases as the planetary radius decreases, in turn increasing the efficiency of heat transport and resulting in lower overall temperature contrasts. While the Coriolis parameter $f=2\Omega \sin{\phi}$ does not depend on the planetary radius $a$, its meridional gradient $\beta=2\Omega\cos{\phi}/a$ is inversely proportional to the planetary radius. The change in the meridional gradient affects the formation and velocity of planetary-scale Rossby waves. The Rossby waves form as a response to potential vorticity gradient, which is caused by differential rotation (different latitudes moving at different velocities). The phase speed of the Rossby waves in the zonal direction is westward relative to the mean zonal flow and directly proportional to the meridional gradient of the Coriolis parameter $\beta$. The phase speed of the Rossby wave relative to the mean zonal flow is $-\beta/(k^2+l^2)$, where $k$ and $l$ are the zonal and meridional wavenumbers, respectively \citep[see, e.g.][]{vallis2006textbook}. Therefore we expect that decreasing the radius would lead to the rise of faster westward Rossby waves compared to the hot Jupiter regime.
   
   The reference geopotential $\overline{\Phi}$ comes from separating the geopotential into the time invariant spatial mean $\overline{\Phi}$ and the time-dependent deviation from the mean $\Phi$ (for a derivation from the continuity and momentum equations, see, e.g. \citet{hack1992description}). The reference geopotential $\overline{\Phi}$ can additionally be thought of as geopotential at scale height $H$, $\overline{\Phi}=gH$, as in \citet{perez2013atmospheric}. We nominally select $g=9.8$ ms$^{-1}$$^2$, but it should be noted that $g$ is included as a factor in geopotential and does not appear separately in the model. Within the context of a shallow-water model, different values of $\overline{\Phi}$ can be interpreted as probing atmospheric layers with different scale heights, which in turn depend on the mean molecular weight of the atmosphere and the average temperature of the layer.  The sample of reference geopotential $\overline{\Phi}$, varying from $4 \times 10^6$ to $1 \times 10^6$ m$^2$s$^{-2}$, is calculated based on a representative sample of equilibrium temperatures ranging from $400$ K to $1200$ K and the mean molecular weights ranging from 2.3 to 5.5.
   This range of mean molecular masses corresponds to atmospheric metallicities ranging from 1 to 200 times solar values \citep[e.g.][]{goyal2019}, which is consistent with the expected range of values for exoplanets with masses between three and thirty Earth masses \citep{wakeford2020exoplanet,welbanks2019}. We give the details of this calculation in Section \ref{sec:phibar-choices}. Our model is agnostic of atmospheric composition, but we have sampled a broad range of values of $\overline{\Phi}$, which incorporate the dependence on atmospheric composition and temperature via the scale height. Further, we have modeled those layers under a variety of forcing conditions, which are related to photospheric opacities and stellar insolation that set photospheric pressures and temperatures.

\begin{figure} 
  \begin{center}
    \includegraphics[width=0.6\textwidth]{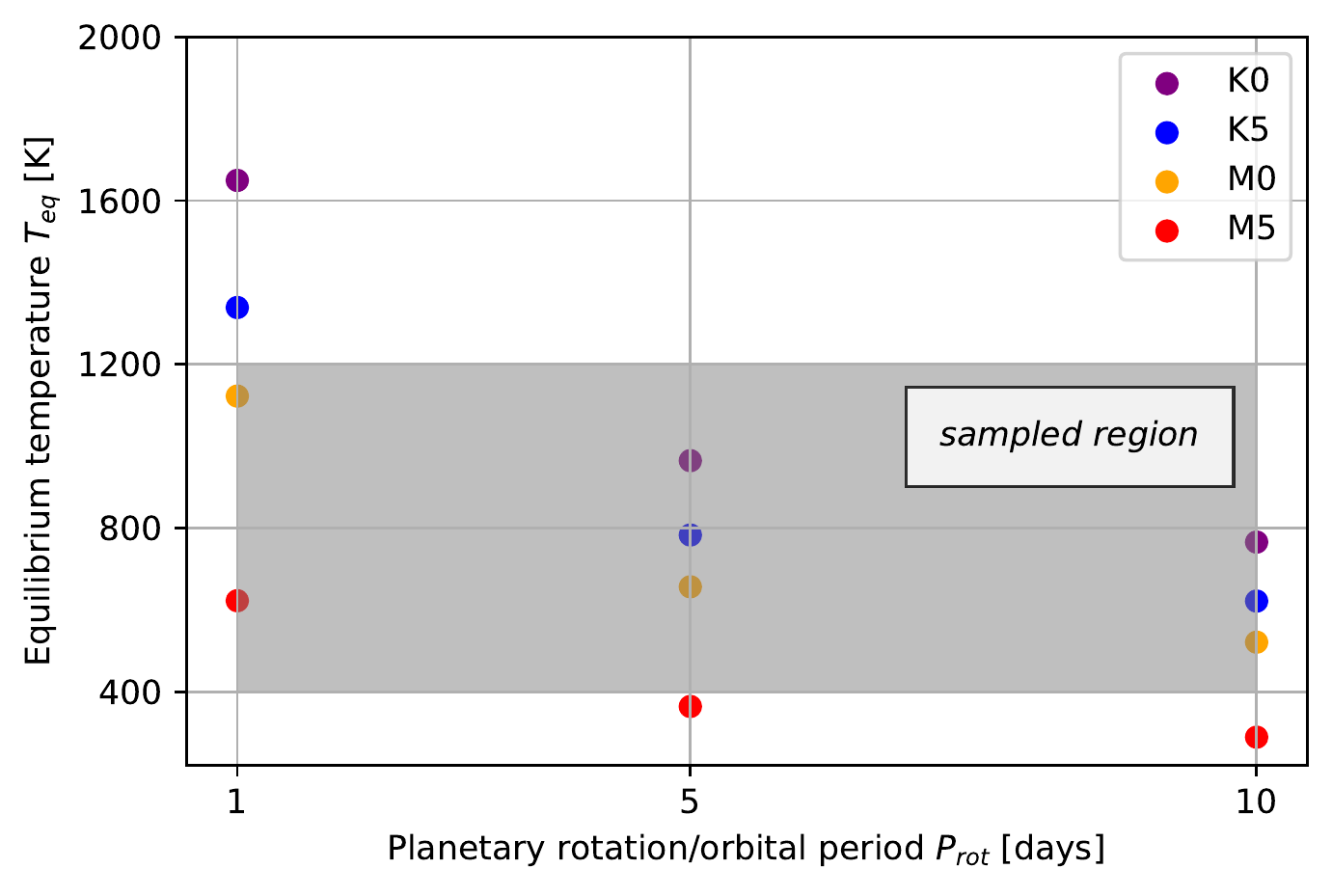}
    \caption{Planetary equilibrium temperature as a function of rotation period for planets around K0 (purple), K5 (blue), M0 (yellow), and M5 (red) stars. The chosen grid of rotation periods of 1, 5, and 10 days and of planetary equilibrium temperatures of 400, 800, and 1200 K  generates a realistic sample. The sample region is highlighted in grey. The stellar parameters were taken from \citet{zombeck1990book}. }
    \label{fig:prot-plot}
  \end{center}
 
\end{figure}

    We vary the planetary rotation period $P_{\rm rot}$ to be 1, 5, and 10 Earth days to reflect a variety of plausible rotation periods of synchronously rotating sub-Neptunes \citep{yang2014strong}. We are assuming $P_{\rm rot}$ equal to the orbital period $P_{\rm orb}$. Figure \ref{fig:prot-plot} illustrates the equilibrium temperatures in the sampled region as a function of planetary rotation periods. All the planets in our sample lie within the tidal locking distance (varying from $0.1$ to $0.8$ AU). Their short orbital periods imply that our simulations could be matched to plausible targets for atmospheric characterization observations with facilities like JWST. Our samples of reference geopotentials and rotation periods fall within a plausible window for a variety of potential insolations.
   
   Gravity is held constant at $g=9.8$ ms$^{-1}$. While this value is based on Earth, there is reason to believe that this gravity value might be representative of the known exoplanet population \citep{ballesteros2016walking}. This value is the reduced gravity, which measures the restoring force at the interface between the two layers. We follow \citet{perez2013atmospheric} in their interpretation of $g$ in the shallow-water context.

   The radiative timescale $\tau_{\rm rad}$  is the time for gas parcels to reach local radiative equilibrium. We follow \citet{showman2002}, where 
 \begin{equation}
    \tau_{\text{rad}}\sim \dfrac{\Delta p}{g}\dfrac{c_{p}}{4 \sigma T^3},
 \end{equation}
 where $\Delta p$ is the thickness of the pressure layer in Pascals, $c_p$ is the specific heat capacity at pressure $p$, $\sigma$ is the Stefan-Boltzmann constant, and $T$ is the temperature of the layer. The radiative timescale represents how quickly the parcels adjust to radiative changes. While $\tau_{\rm rad}$ is a free parameter in this study, varying the radiative timescale between 0.1, 1, and 10 Earth days corresponds to probing a variety of pressure level and temperature combinations. While the radiative time constant $\tau_{\rm rad}$ decreases sharply with increasing temperature, note that the estimate breaks down in deep, optically thick atmospheres, e.g. deep inside hot Jupiter atmospheres, $\tau_{\rm rad}$ is long. Our samples span the range from a few hours, typical for hot Jupiters \citep{showman2010atmospheric} to a few days, approximately corresponding to Earth \citep{showman2008b}. \citet{showman2002} predicted that on hot Jupiters, large day/night fractional differences will persist when the radiative timescale $\tau_{\rm rad}$ is much shorter than the advective timescale  $\tau_{\rm adv}$, and small day/night fractional differences will appear when the radiative timescale is much longer than the advective timescale. Our sample of radiative timescales allows us to explore the relationship between radiative timescale and day/night contrast in the context of sub-Neptunes.
 
  We investigate three different relative radiative forcing amplitudes, which prescribe the radiative equilibrium temperature differences between the dayside and the nightside. We let $\Delta \Phi_{\rm eq}/\overline{\Phi}=1$, 0.5, and 0.1, corresponding to hot, warm, and cool sub-Neptune planets.  In the context of hot Jupiters, the high-contrast regime has been explored by \citet{perez2013atmospheric}, and the medium-contrast regime has been explored by \citet{liu2013initialconditions}. Both of the above studies also include a low-contrast regime, with $\Delta \Phi_{\rm eq}/\overline{\Phi}$ varying from as low as $0.001$ to $0.1$, corresponding to the lowest forcing amplitude in our study. In conjunction with our choices of reference geopotential $\overline{\Phi}$, the choice of three forcing amplitudes allows us to span the vast parameter space of potential stellar forcings and atmospheric scale heights with our simplified model.

   \subsection{Motivation for the choices of $\overline{\Phi}$}
   \label{sec:phibar-choices}
 We have chosen the mean geopotential $\overline{\Phi}$ by calculating the scale heights for a variety of equilibrium temperatures and planetary atmospheric compositions. While the shallow-water model is agnostic about atmospheric compositions, we have selected the sampled scale heights and forcing amplitudes to encompass a variety of plausible scenarios. 
 We select the mean geopotential to correspond to scale height $H$, which we compute as follows:
 \begin{equation}
     \overline{\Phi}=gH=\dfrac{kT_{\rm eq}}{m},
 \end{equation}
 
 where $k$ is the Boltzmann constant, $T_{\rm eq}$ is the equilibrium temperature, and $m$ (kg/molecule) is the mean molar mass of the atmospheric constituents.  The range of scale heights was computed using temperatures in the $400-1200$ K range and mean molecular weights of 2.3-5.5 times the mass of a hydrogen atom, corresponding to 1-200 multiples of solar metallicity. The resulting values for $\overline{\Phi}$ were relatively evenly spread out between $6\times 10^5$ m$^2$s$^{-2}$ and $4.3 \times 10^6$ m$^2$s$^{-2}$, so the samples of $10^6$ m$^2$s$^{-2}$, $2\times 10^6$m$^2$s$^{-2}$, $3\times 10^6$ m$^2$s$^{-2}$, and $4 \times 10^6$ m$^2$s$^{-2}$ efficiently sample the parameter space, despite combining two parameters into one. The models at high $\overline{\Phi}$ can be interpreted as corresponding to planets with a hydrogen-dominated atmosphere, similar to that of Jupiter. A decrease in $\overline{\Phi}$ can be thought of as a transition to a more metal-rich atmosphere. While the strength of stellar forcing is not independent of reference geopotential in our model, our simulations span three forcing regimes for each scale height, allowing us to explore the effects of scale height in the context of the shallow-water model.

\section{Results} \label{sec:results}
\begin{table}[]
\begin{tabular}{|llll|}
\hline
\multicolumn{4}{|c|}{Strong forcing}                                                                                                                                                                                                                                                                                                                                                                                                                                                                                         \\ \hline
\multicolumn{1}{|l|}{}                      & \multicolumn{1}{l|}{$\tau_{\rm rad}=0.1$ days}                                                                                                                                      & \multicolumn{1}{l|}{$\tau_{\rm rad}=1$ day}                                                                                                          & $\tau_{\rm rad}=10$ days                                                                                                        \\ \hline
\multicolumn{1}{|l|}{$P_{\rm rot}=1$ day}   & \multicolumn{1}{l|}{\begin{tabular}[c]{@{}l@{}}$\mathbf{V}_{\rm max}\approx 1700$ ms$^{-1}$\\ $\Delta\Phi \approx 1.1\times 10^6$ m$^2$/s$^2$ \\Ro=0.62\end{tabular}}                               & \multicolumn{1}{l|}{\begin{tabular}[c]{@{}l@{}}$\mathbf{V}_{\rm max}\approx 300$ ms$^{-1}$\\ $\Delta\Phi \approx 5.1\times 10^3$ m$^2$/s$^2$ \\Ro=0.11 \end{tabular}} & \begin{tabular}[c]{@{}l@{}}$\mathbf{V}_{\rm max}\approx 160$ ms$^{-1}$\\ $\Delta\Phi \approx 3.4\times 10^2$ m$^2$/s$^2$\\Ro=0.06\end{tabular} \\ \hline
\multicolumn{1}{|l|}{$P_{\rm rot}=5$ days}  & \multicolumn{1}{l|}{\begin{tabular}[c]{@{}l@{}}$\mathbf{V}_{\rm max}\approx 900$ ms$^{-1}$\\ $\Delta\Phi \approx 5.7\times 10^5$ m$^2$/s$^2$\\Ro=1.59\end{tabular}}                                & \multicolumn{1}{l|}{\begin{tabular}[c]{@{}l@{}}$\mathbf{V}_{\rm max}\approx 570$ ms$^{-1}$\\ $\Delta\Phi \approx 2.4\times 10^5$ m$^2$/s$^2$\\Ro=1.03\end{tabular}} & \begin{tabular}[c]{@{}l@{}}$\mathbf{V}_{\rm max}\approx 80$ ms$^{-1}$\\ $\Delta\Phi \approx 8.4\times 10^2$ m$^2$/s$^2$\\Ro=0.15\end{tabular}  \\ \hline
\multicolumn{1}{|l|}{$P_{\rm rot}=10$ days} & \multicolumn{1}{l|}{\begin{tabular}[c]{@{}l@{}}$\mathbf{V}_{\rm max}\approx 600$ ms$^{-1}$\\ $U_{\rm max}\approx 520$ ms$^{-1}$\\ $\Delta\Phi \approx 7.2\times 10^3$ m$^2$/s$^2$\\Ro=1.87\end{tabular}} & \multicolumn{1}{l|}{\begin{tabular}[c]{@{}l@{}}$\mathbf{V}_{\rm max}\approx 330$ ms$^{-1}$\\ $\Delta\Phi \approx 1.1\times 10^5$ m$^2$/s$^2$\\Ro=1.17\end{tabular}} & \begin{tabular}[c]{@{}l@{}}$\mathbf{V}_{\rm max}\approx 145$ ms$^{-1}$\\ $\Delta\Phi \approx 7.1\times 10^3$ m$^2$/s$^2$\\Ro=0.53\end{tabular} \\ \hline
\multicolumn{4}{|c|}{Medium Forcing}                                                                                                                                                                                                                                                                                                                                                                                                                                                                                       \\ \hline
\multicolumn{1}{|l|}{}                      & \multicolumn{1}{l|}{$\tau_{\rm rad}=0.1$ days}                                                                                                                                      & \multicolumn{1}{l|}{$\tau_{\rm rad}=1$ day}                                                                                                          & $\tau_{\rm rad}=10$ days                                                                                                        \\ \hline
\multicolumn{1}{|l|}{$P_{\rm rot}=1$ day}   & \multicolumn{1}{l|}{\begin{tabular}[c]{@{}l@{}}$\mathbf{V}_{\rm max}\approx 1200$ ms$^{-1}$\\ $\Delta\Phi \approx 6.4\times 10^5$ m$^2$/s$^2$\\Ro=0.44\end{tabular}}                               & \multicolumn{1}{l|}{\begin{tabular}[c]{@{}l@{}}$\mathbf{V}_{\rm max}\approx 160$ ms$^{-1}$\\ $\Delta\Phi \approx 7.1\times 10^3$ m$^2$/s$^2$\\Ro=0.06\end{tabular}} & \begin{tabular}[c]{@{}l@{}}$\mathbf{V}_{\rm max}\approx 120$ ms$^{-1}$\\ $\Delta\Phi \approx 3.4\times 10^2$ m$^2$/s$^2$\\Ro=0.04\end{tabular} \\ \hline
\multicolumn{1}{|l|}{$P_{\rm rot}=5$ days}  & \multicolumn{1}{l|}{\begin{tabular}[c]{@{}l@{}}$\mathbf{V}_{\rm max}\approx 600$ ms$^{-1}$\\ $U_{\rm max}\approx 675$ ms$^{-1}$\\ $\Delta\Phi \approx 7.2\times 10^4$ m$^2$/s$^2$\\Ro=1.42\end{tabular}} & \multicolumn{1}{l|}{\begin{tabular}[c]{@{}l@{}}$\mathbf{V}_{\rm max}\approx 490$ ms$^{-1}$\\ $\Delta\Phi \approx 2.4\times 10^4$ m$^2$/s$^2$\\Ro=0.87\end{tabular}} & \begin{tabular}[c]{@{}l@{}}$\mathbf{V}_{\rm max}\approx 80$ ms$^{-1}$\\ $\Delta\Phi \approx 2.0\times 10^2$ m$^2$/s$^2$\\Ro=0.15\end{tabular}  \\ \hline
\multicolumn{1}{|l|}{$P_{\rm rot}=10$ days} & \multicolumn{1}{l|}{\begin{tabular}[c]{@{}l@{}}$\mathbf{V}_{\rm max}\approx 430$ ms$^{-1}$\\ $\Delta\Phi \approx 1.3\times 10^4$ m$^2$/s$^2$\\Ro=1.53\end{tabular}}                                & \multicolumn{1}{l|}{\begin{tabular}[c]{@{}l@{}}$\mathbf{V}_{\rm max}\approx 280$ ms$^{-1}$\\ $\Delta\Phi \approx 8.7\times 10^3$ m$^2$/s$^2$\\Ro=0.99\end{tabular}} & \begin{tabular}[c]{@{}l@{}}$\mathbf{V}_{\rm max}\approx 200$ ms$^{-1}$\\ $\Delta\Phi \approx 1.5\times 10^4$ m$^2$/s$^2$\\Ro=0.71\end{tabular} \\ \hline
\multicolumn{4}{|c|}{Weak Forcing}                                                                                                                                                                                                                                                                                                                                                                                                                                                                                          \\ \hline
\multicolumn{1}{|l|}{}                      & \multicolumn{1}{l|}{$\tau_{\rm rad}=0.1$ days}                                                                                                                                      & \multicolumn{1}{l|}{$\tau_{\rm rad}=1$ day}                                                                                                          & $\tau_{\rm rad}=10$ days                                                                                                        \\ \hline
\multicolumn{1}{|l|}{$P_{\rm rot}=1$ day}   & \multicolumn{1}{l|}{\begin{tabular}[c]{@{}l@{}}$\mathbf{V}_{\rm max}\approx 330$ ms$^{-1}$\\ $\Delta\Phi \approx 4.1\times 10^4$ m$^2$/s$^2$\\Ro=0.10\end{tabular}}                                & \multicolumn{1}{l|}{\begin{tabular}[c]{@{}l@{}}$\mathbf{V}_{\rm max}\approx 50$ ms$^{-1}$\\ $\Delta\Phi \approx 1.4\times 10^3$ m$^2$/s$^2$\\Ro=0.02\end{tabular}}  & \begin{tabular}[c]{@{}l@{}}$\mathbf{V}_{\rm max}\approx 60$ ms$^{-1}$\\ $\Delta\Phi \approx 6\times 10$ m$^2$/s$^2$\\Ro=0.02\end{tabular}      \\ \hline
\multicolumn{1}{|l|}{$P_{\rm rot}=5$ days}  & \multicolumn{1}{l|}{\begin{tabular}[c]{@{}l@{}}$\mathbf{V}_{\rm max}\approx 440$ ms$^{-1}$\\ $\Delta\Phi \approx 5.3\times 10^4$ m$^2$/s$^2$\\Ro=0.79\end{tabular}}                                & \multicolumn{1}{l|}{\begin{tabular}[c]{@{}l@{}}$\mathbf{V}_{\rm max}\approx 80$ ms$^{-1}$\\ $\Delta\Phi \approx -5\times 10$ m$^2$/s$^2$\\Ro=0.16\end{tabular}}     & \begin{tabular}[c]{@{}l@{}}$\mathbf{V}_{\rm max}\approx 100$ ms$^{-1}$\\ $\Delta\Phi \approx 2.4\times 10^2$ m$^2$/s$^2$\\Ro=0.20\end{tabular} \\ \hline
\multicolumn{1}{|l|}{$P_{\rm rot}=10$ days} & \multicolumn{1}{l|}{\begin{tabular}[c]{@{}l@{}}$\mathbf{V}_{\rm max}\approx 310$ ms$^{-1}$\\ $\Delta\Phi \approx 1.4\times 10^4$ m$^2$/s$^2$\\Ro=1.13\end{tabular}}                                & \multicolumn{1}{l|}{\begin{tabular}[c]{@{}l@{}}$\mathbf{V}_{\rm max}\approx 160$ ms$^{-1}$\\ $\Delta\Phi \approx 8.0\times 10^3$ m$^2$/s$^2$\\Ro=0.82\end{tabular}} & \begin{tabular}[c]{@{}l@{}}$\mathbf{V}_{\rm max}\approx 100$ ms$^{-1}$\\ $\Delta\Phi \approx 6\times 10$ m$^2$/s$^2$\\Ro=0.37\end{tabular}     \\ \hline
\end{tabular}

\caption{Typical maximal wind speeds, day/night geopotential contrast, and Rossby number for a representative scale height $\overline{\Phi}=4\times 10^6$ m$^2$/s$^2$ based on temporal averages. The geopotential contrast is computed by subtracting the nightside average from the dayside average. When maximal wind speeds $\mathbf{V}_{\rm max}$ values differ from maximal values of the zonal component $U_{\rm max}$,  the zonal maximum $U_{\rm max}$ is also included.}
\label{table:phibar-4-summary}
\end{table}

\subsection{Extension to Sub-Neptunes}
\label{sec:turn-down-radius}

\begin{figure}
  \centering
  \includegraphics[width=0.8\linewidth]{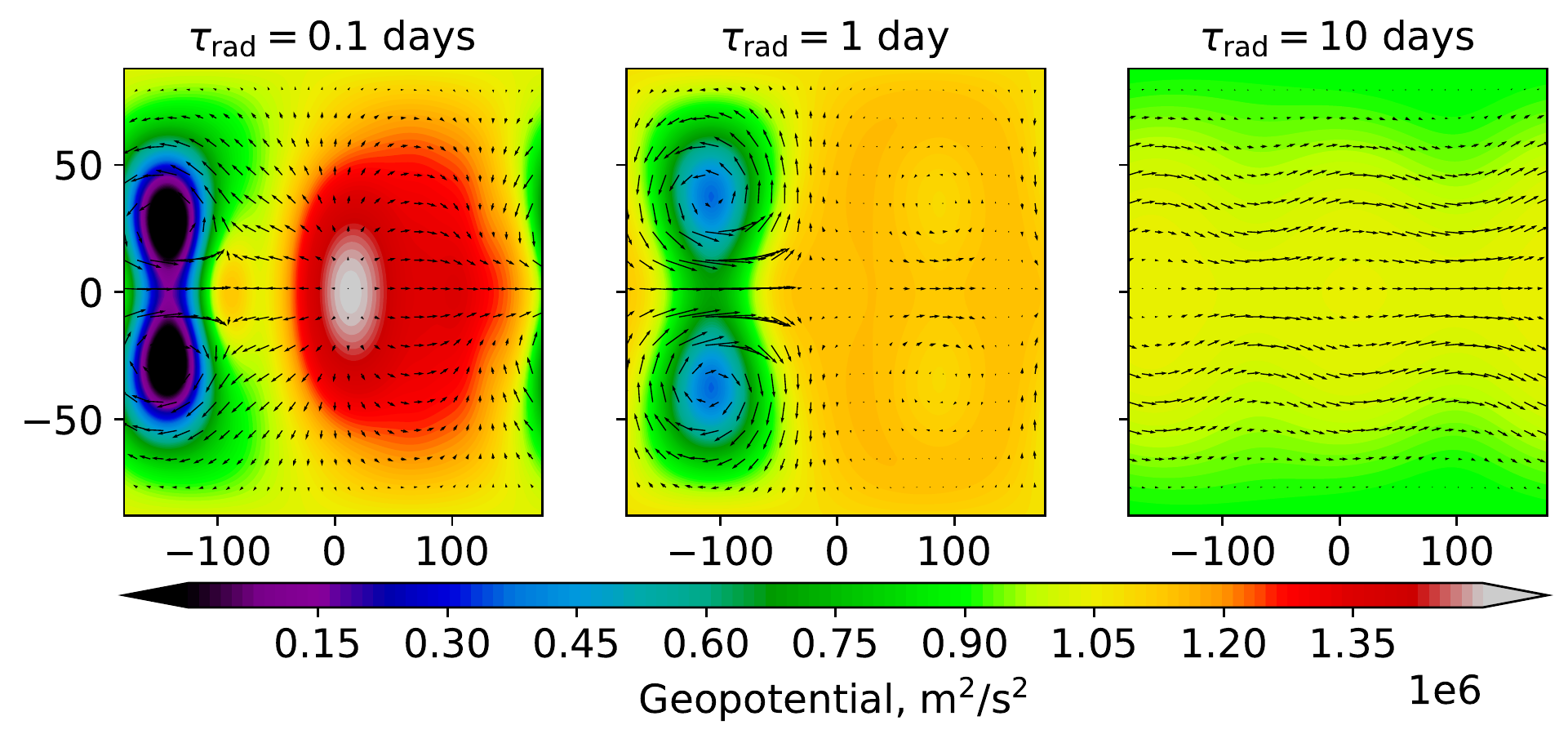}
  \caption{Recreation of a hot Jupiter regime in Figure 3 in \cite{perez2013atmospheric}, with planetary radius turned down to $a=3a_{\oplus}=1.3\times10^7$ m. Shown are the geopotential contour maps and the wind vector fields. The remaining planetary parameters are as follows: planetary rotation rate $\Omega=3.2\times10^{-5}$ rad/s (equivalent to $P_{\rm rot}=2.27$ days), no drag, reference geopotential $\overline{\Phi}=4 \times 10^6$, for strong radiative forcing ($\Delta \Phi_{\rm eq}/\overline{\Phi}=1$).  Three radiative time scales are depicted: $\tau_{\rm rad=0.1}$ days (left panel), $\tau_{\rm rad=1}$ day (center panel), and $\tau_{\rm rad=10}$ days (right panel). We have subtracted the constant value of $\overline{\Phi}$ from each panel. Panels share the same color scale for geopotential, but winds are normalized independently in each panel. The substellar point is located at $(0^{\circ},0^{\circ})$ for each panel.}
  \label{sub-Neptune-PBS}
\end{figure}

    After benchmarking our model in the hot Jupiter regime, we transition the model toward a regime appropriate for sub-Neptunes by first decreasing the planetary radius $a$. We decrease the planetary radius from that of Jupiter to equal three times Earth radius ($a=1.9\times 10^7$ m) while keeping the rest of the parameters identical to our validation regime: a hot Jupiter based on HD~189733b, explored in Section \ref{sec:validation}. This change bridges the previously explored hot Jupiter regimes and the sub-Neptune regimes, which are the focus of this work. Compared to our replication of a hot Jupiter regime (shown in Figure \ref{PBS-recreation}), reducing the planetary radius preserves the transition from day-to-night flow for short radiative timescales ($0.1$ days) to a jet-dominated flow for long radiative timescales ($10$ days). However, the sub-Neptune simulations shown in Figure \ref{sub-Neptune-PBS} exhibit lower geopotential height contrast both in the zonal direction (between the dayside and the nightside) and in the meridional direction (between the equator and the pole) than the hot-Jupiter simulations. We also observe that for short and medium values of $\tau_{\rm rad}$ (0.1 and 1 days, left and center panels, respectively), the nightside cyclonic vortices are more confined in latitude. For the short timescale $\tau_{\rm rad}=0.1$ days, the vortices extend from $\approx10^\circ$ to $\approx 50^\circ$ latitude, with wind speeds $|\mathbf{V}|\approx10^3$ ms$^{-1}$ and the geopotential contrast of $\approx10^6$ within the vortices. For the medium radiative timescale $\tau_{\rm rad}=1$ day, the vortices extend from $\approx15^\circ$ to $\approx 70^\circ$ latitude, with wind speeds $|\mathbf{V}|\approx8\times10^2$ ms$^{-1}$ and geopotential contrast $\approx6\times10^5$ within the vortices. While the rotation period $P_{\rm rot}$ was not changed from that one used in \citet{perez2013atmospheric}, the change in planetary radius leads to an inversely proportional change in the beta parameter $\beta=\frac{\partial f}{\partial y}=2 \Omega \cos \phi_0/a$. The beta parameter is in turn proportional to the speed of Rossby waves. In the left panel, for $\tau_{\rm rad}=0.1$ days, the vortices are centered at $\sim 30^\circ$ latitude.  In the center panel, where $\tau_{\rm rad}=1$ day, the vortices are larger and father from the equator, at $\sim40^\circ$ latitude, but exhibit lower geopotential contrast. The center panel simulation also exhibits a secondary, warmer pair of cyclonic vortices. This behavior is qualitatively different from the $\tau_{\rm rad}=1$ day regime in Figure \ref{PBS-recreation}, where anticyclonic vortices appear on the dayside. In the right panel, where the radiative timescale $\tau_{\rm rad}=10$ days, the simulation exhibits less equator-to-pole variation than the equivalent regime in Figure \ref{PBS-recreation}. Reducing the planetary radius by $75\%$ has led to the emergence of significant changes in global scale atmospheric flow patterns and wind speeds, showing distinct differences in global circulation patterns between hot Jupiters and sub-Neptunes.
    
    We now proceed to model sub-Neptunes by performing a parameter sweep of rotation periods $P_{\rm rot}$ and radiative timescales $\tau_{\rm rad}$ in three forcing regimes: strong forcing ($\Delta\Phi_{\rm eq}/\overline{\Phi}=1$, Section \ref{sec:strong-forcing}), medium forcing ($\Delta\Phi_{\rm eq}/\overline{\Phi}=0.5$), and weak forcing ($\Delta\Phi_{\rm eq}/\overline{\Phi}=0.1$). The effects in response to changing forcing strength are discussed in Section \ref{sec:response}. In Table \ref{table:phibar-4-summary}, we summarize the maximal wind speeds $\mathbf{V}_{\rm max}$, and the geopotential contrast between the dayside and the nightside. The maximal equatorial zonal speeds $U_{\rm max}$ are similar to the maximal overall wind speeds. While our parameter sweep includes four scale height values, a representative scale height corresponding to $\overline{\Phi}=4 \times 10^6$ was chosen to be represented in the table.

  \subsection{Strong forcing simulations}
  \label{sec:strong-forcing}

     In this section, we explore what we are calling ``strong forcing'' simulations in the context of our ensemble. For these simulations the prescribed local radiative equilibrium day-night forcing contrast is equal to the the nightside geopotential at radiative equilibrium ($\Delta \Phi_{\rm eq}/\overline{\Phi}=1$). For strong forcing regimes, the equilibrium day-night contrast is of the same magnitude as the equilibrium nightside geopotential $\overline{\Phi}$. Note that changing the scale height results in changing $\Delta \Phi_{\rm eq}$, and so in our formulation, the stellar forcing depends on the value of $\overline{\Phi}$.

\begin{figure}
  \centering
  \includegraphics[width=0.9\linewidth]{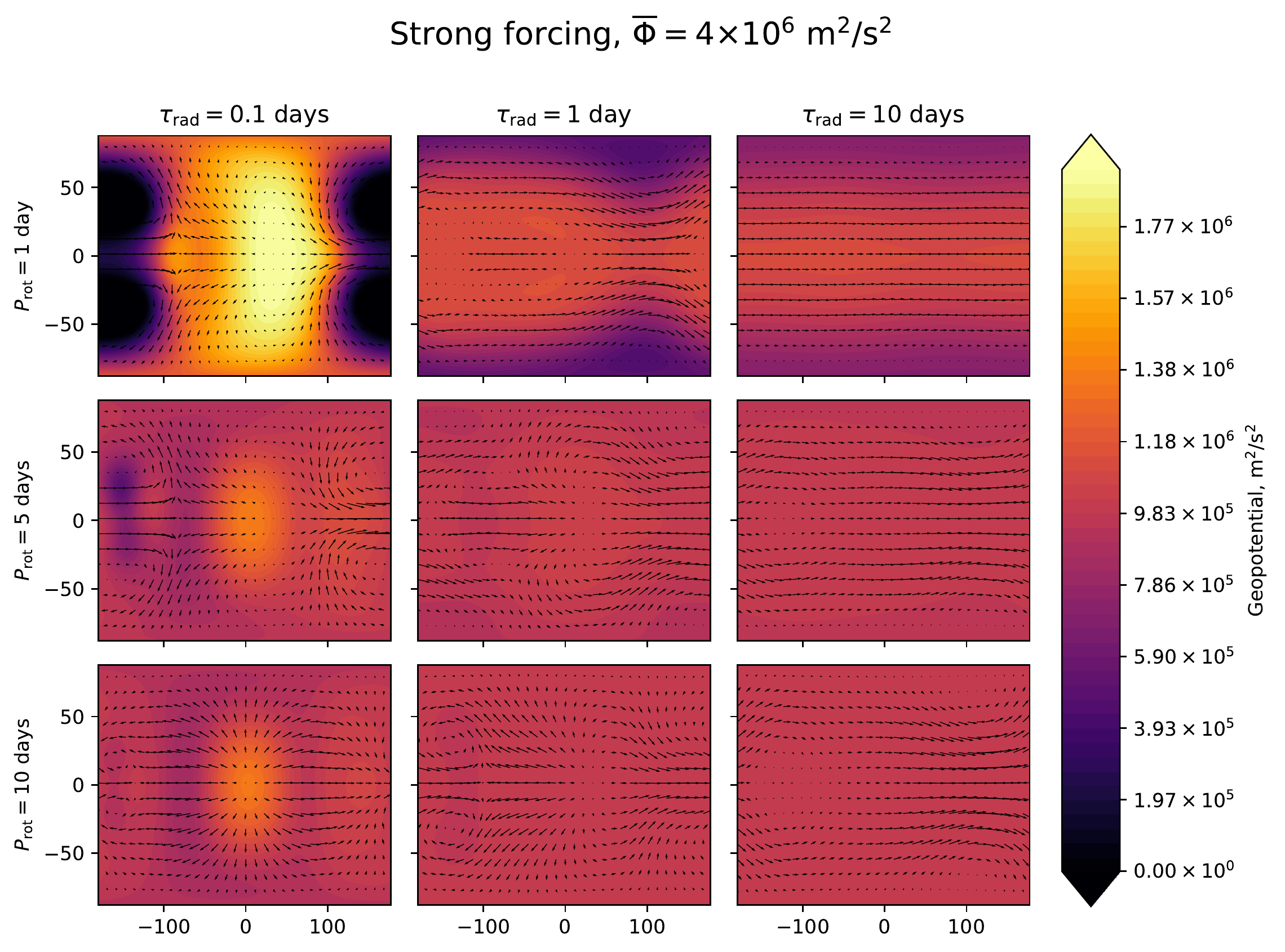}
  \caption{Latitude-longitude maps of the steady-state geopotential anomaly and wind vector fields for the strong forcing regime ($\Delta \Phi_{\rm eq}/\overline{\Phi}=1$) at reference geopotential $\overline{\Phi}=4 \times 10^6$ m$^2$s$^{-2}$, averaged over the last 100 days of the simulation. The constant value of $\overline{\Phi}$ has been subtracted from each panel. Each panel in the grid corresponds to a unique combination of planetary rotation period $P_{\rm rot}$ and radiative timescale $\tau_{\rm rad}$. Panels share the same color scale for geopotential, but winds are normalized independently in each panel. The substellar point is located at $(0^{\circ},0^{\circ})$ for each panel.}
  \label{high-Phibar-4}
\end{figure}

\begin{figure}
  \centering
  \includegraphics[width=0.9\linewidth]{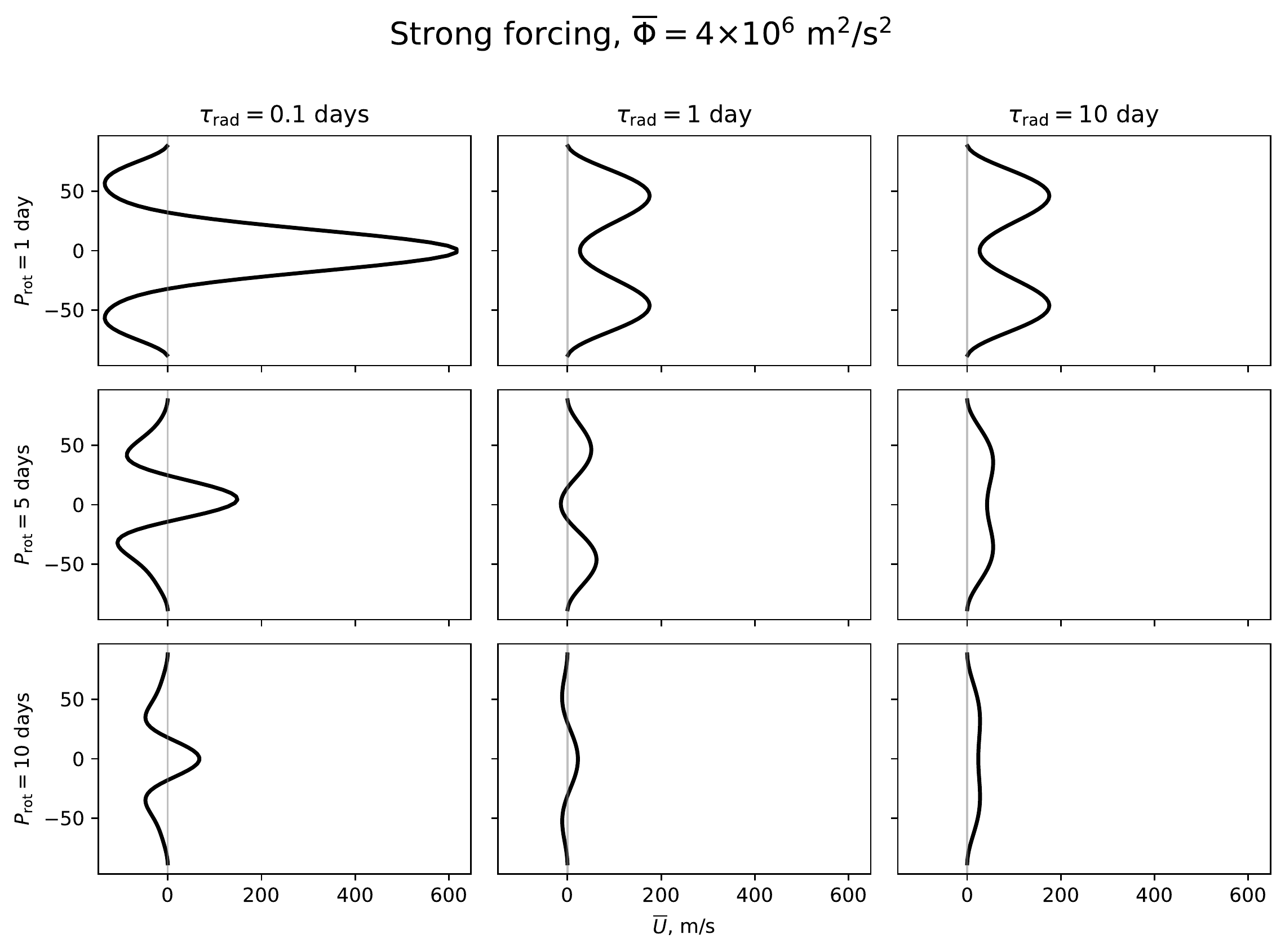}
  \caption{Mean-zonal winds $\overline{U}$ for the steady-state solution for the strong forcing regime ($\Delta \Phi_{\rm eq}/\overline{\Phi}=1$) at reference geopotential $\overline{\Phi}=4 \times 10^6$ m$^2$s$^{-2}$, averaged over the last 100 days of the simulation.  Each panel in the grid corresponds to a unique combination of planetary rotation period $P_{\rm rot}$ and radiative timescale $\tau_{\rm rad}$.}
  \label{high-Phibar-4-wind}
\end{figure}

     First we consider $\overline{\Phi}=4 \times 10^6$ m$^2$s$^{-2}$, corresponding to a large scale height. While our simplified shallow-water model is agnostic to the exact specification of the planetary composition, a large scale height could correspond, for example, to a predominantly H/He atmosphere with $\approx1\times$ solar metallicity. The resulting geopotential profile for this regime is shown in Figure \ref{high-Phibar-4}. The corresponding mean zonal wind profiles are shown in Figure \ref{high-Phibar-4-wind}. The shortest-timescale regime (with radiative timescale $\tau_{\rm rad}=0.1$ days and rotation period $P_{\rm rot}=1$ day) exhibits the highest day-to night contrast, highest equator-to-pole contrast, and highest wind speeds. The longest-timescale regime (with radiative timescale $\tau_{\rm rad}=10$ days and rotation period $P_{\rm rot}=10$ days) exhibits the lowest day-to night contrast, lowest equator-to-pole contrast, and lowest wind speeds. We discuss these transitions in more detail in the following paragraphs.
     
      In the short-timescales regime shown in the top left panel of Figure \ref{high-Phibar-4}, where radiative timescale $\tau_{\rm rad}=0.1$ days and rotation period $P_{\rm rot}=1$ day, strong forcing is combined with strong rotational effects. This regime exhibits high day-night contrast combined with a high equator-to-pole contrast. We observe atmospheric superrotation with a strong eastward equatorial jet (the mean zonal wind speeds reach $\sim0.6$ km s$^{-1}$ at the equator) that shifts the dayside hotspot to the east. A combination of eastward Kelvin waves and westward Rossby waves results in equatorial superrotation that gives rise to a hotspot shape and offset from the substellar longitude characteristic for hot Jupiters \citep{showman2020atmospheric}. Indeed, this strongly forced, short radiative timescale regime is closest to a hot Jupiter regime among the ones we explore. While on hot Jupiters, the equatorial winds tend to be uniformly eastward, the simulation exhibits some regions of equatorial westward velocities, which are partly driven by the relative strength of the Coriolis force, which is higher for a smaller planet with $P_{\rm rot}=1$ day. On the night side, the simulation exhibits stationary polar vortices. 
      
      In contrast, in the right column of Figure \ref{high-Phibar-4}, long radiative timescale $\tau_{\rm rad}=10$ days results in the formation of a jet-dominated pattern with little longitudinal variation, regardless of the rotation period. The equator-to-pole variation is significantly lower than that of the prescribed radiative equilibrium. No vortices are present. For short rotation period $P_{\rm rot}=1$, this regime exhibits an oscillation between a hot equatorial band and two hot tropical bands. While this oscillation is not visible in Figure  \ref{high-Phibar-4}, we discuss it in more detail in Section \ref{sec:tropical-oscillation}.
     
    When the radiative timescale is at its intermediate value in our ensemble, $\tau_{\rm rad}=1$ day, we observe the transition between the regime dominated by the day-night flow, which occurs for shorter radiative timescales, and the one dominated by the jet flow, which occurs for longer radiative timescales.
    In the top center panel of Figure \ref{high-Phibar-4} ($\tau_{\rm rad}=1$ day, $P_{\rm rot}=1$ day), there is some longitudinal variation, but the hottest parts of the planet are no longer confined to the dayside. Polar vortices are present, extending to $\approx 60^\circ$ latitude, centered at the terminator $90^\circ$ east of the substellar point. While there is high equator-to-pole variation, the equator-to-pole contrast is significantly lower than that of the prescribed radiative equilibrium. The average zonal wind speeds exhibit peaks of $\approx 150$ ms$^{-1}$ at $\sim 40^\circ-50^\circ$ latitude, and these wind speeds are much lower than the $\approx 600$ ms$^{-1}$ eastward equatorial jet exhibited by the $\tau_{\rm rad}=0.1$ days regime.
      
    As $P_{\rm rot}$ increases from 1 to 5 days, the eastward equatorial jet becomes weaker, with lower associated mean zonal wind speeds. The models shown in the middle row of Figure \ref{high-Phibar-4} exhibit a robust day-to-night flow pattern. For $\tau_{\rm rad}=0.1$ days (left column), compared to simulations with a lower rotation period, the night-side cyclonic vortices are smaller and closer to the equator (centered at around $\sim 20^\circ-30^\circ$ latitude). The cyclonic vortices correspond to the westward mean zonal winds around those latitudes. The equatorial zonal winds are eastward, and the amplitude is lower than that for shorter rotation period values. In this regime, we observe some asymmetry around the equator, with the cyclonic vortex in the northern hemisphere being larger than in the southern one. The presence of momentum advection across the equator in the presence of symmetrical forcing for a zero-obliquity planet seems to indicate that a pitchfork bifurcation is taking place. We discuss this phenomenon in more detail in Section \ref{sec:pitchfork}.
    
    When $P_{\rm rot}=5$ days and $\tau_{\rm rad}=1$ day, in the center panel of Figure \ref{high-Phibar-4}, we observe some time variability. There is a pair of nightside cyclonic vortices, which migrate in both north-south and east-west directions (although always confined to the nightside) in an oscillation with a period of $\approx25$ days.
     Further increasing the rotation period $P_{\rm rot}$ to 10 days, in the bottom center panel, the simulation exhibits migrating nightside cyclonic vortices with a period of $\approx25$ days. Compared to shorter $P_{\rm rot}$, the equator-to-pole contrast is lower. The presence of time variability and momentum advection across the equator in these regimes emerges in a particular regime of radiative forcing strength and rotation period. While high variability is generally associated with longer timescales, it should be noted that the longest-timescales regime ($P_{\rm rot}=10$ days and $\tau_{\rm rad}=10$ days) exhibits no time variability. 
    
     Overall in the strong-forcing, high-scale-height regime, the shift from short radiative timescales to longer ones corresponds to the shift from a day-to-night flow with nightside vortices to a jet-dominated flow, as well as from a low time-variability to a higher time-variability. Increasing the rotation period typically decreases the equator-to-pole contrast. The shift from $P_{\rm rot}=1$ day to $P_{\rm rot}=5$ days creates a more dramatic change in the atmospheric circulation patterns than the shift from $P_{\rm rot}=5$ days to $P_{\rm rot}=10$ days. 
     
    As expected, the wind speeds are strongest ($\approx 600$ ms$^{-1}$) when both radiative and rotational timescales are short, and weakest ($\approx 10-20$ ms$^{-1}$) when both timescales are longest. Most of our simulations in the high-forcing regime exhibit an equatorial jet pattern. However, when the radiative timescale is long and the rotation period is short (top center and top right in Figure \ref{high-Phibar-4-wind}), the average zonal wind speeds exhibit two tropical peaks. We discuss this transition in more detail in Section \ref{sec:disc-wind-trends}.

    \subsection{Response to changing scale height and forcing strength}
    \label{sec:response}

\begin{figure}
      \centering
      \includegraphics[width=0.8\linewidth]{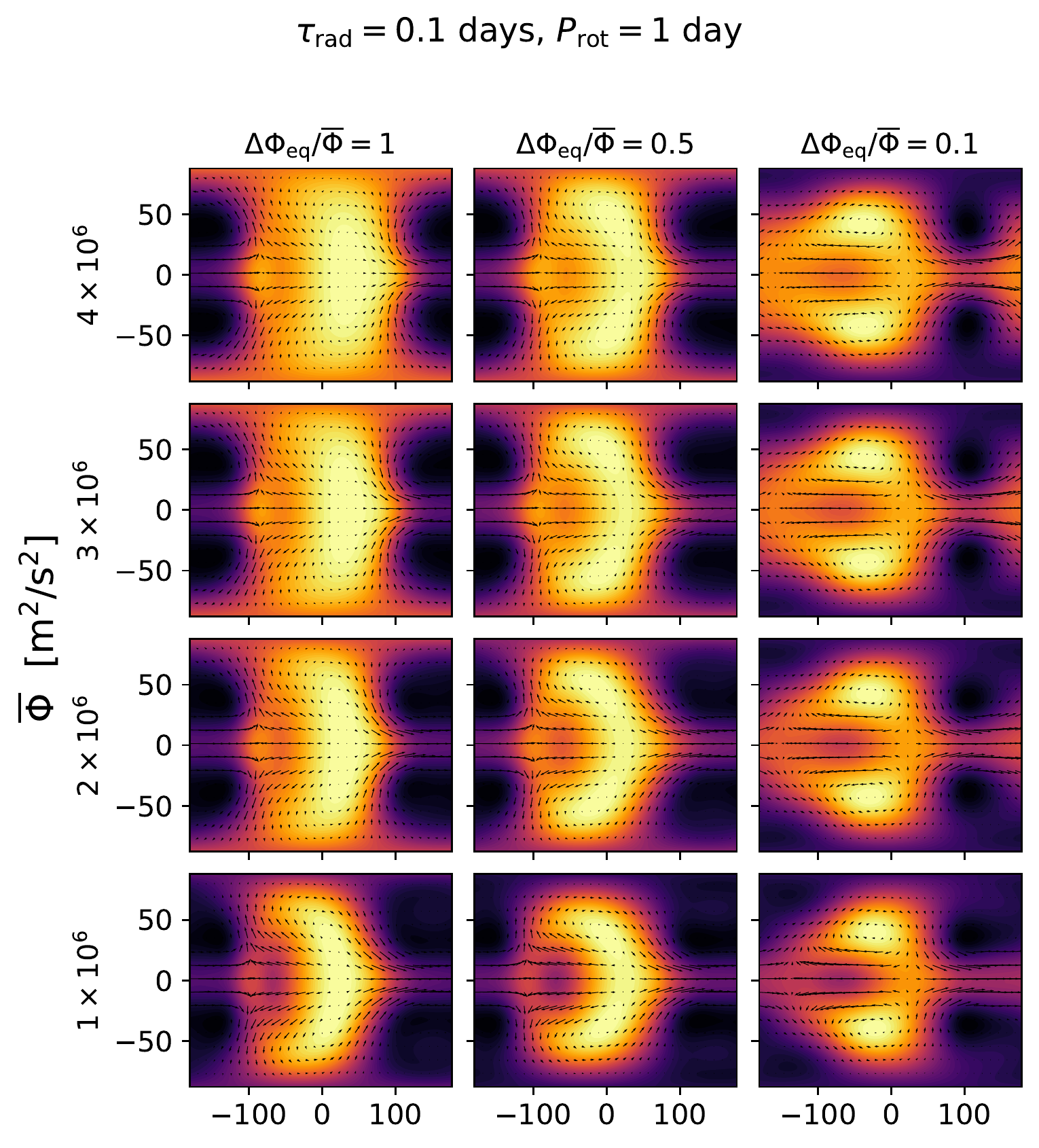}
      \caption{Latitude-longitude maps of the steady-state geopotential anomaly and wind vector fields for the shortest timescales regime (radiative timescale $\tau_{\rm rad}=0.1$ days, rotation period $P_{\rm rot}=1$ day), averaged over the last 100 days of the simulation. Each panel in the grid corresponds to a unique combination of forcing strength $\Delta \Phi_{\rm eq}/\overline{\Phi}$ and reference geopotential $\overline{\Phi}$. The wind vectors are normalized independently in each panel. The substellar point is located at $(0^{\circ},0^{\circ})$ for each panel. The colorbar limits have been adjusted individually for each panel to ease comparison of atmospheric patterns across orders of magnitude.}
      \label{fig:geopot_grid_fig}
    \end{figure}

\begin{figure}
      \centering
      \includegraphics[width=0.8\linewidth]{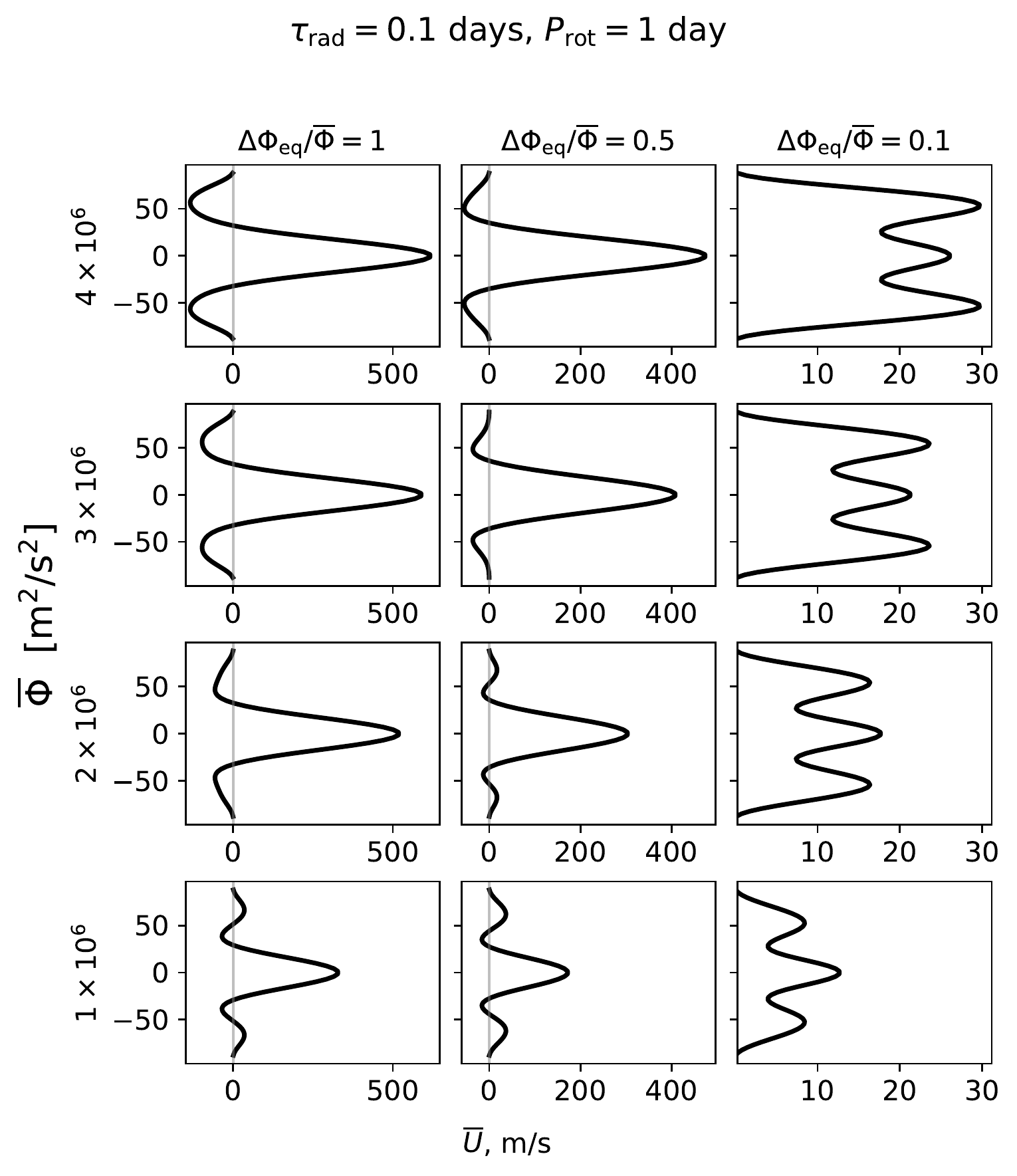}
      \caption{ {Mean-zonal winds $\overline{U}$ for for the shortest timescales regime (radiative timescale $\tau_{\rm rad}=0.1$ days, rotation period $P_{\rm rot}=1$ day), averaged over the last 100 days of the simulation. Each panel in the grid corresponds to a unique combination of forcing strength $\Delta \Phi_{\rm eq}/\overline{\Phi}$ and reference geopotential $\overline{\Phi}$. Note that the axis limits have been adjusted individually for each column in order to highlight the features for the smaller wind speeds that arise in the weak-forcing regime.}}
      \label{fig:wind_grid_fig}
    \end{figure}

    In this section, we present the trends in our simulations as we vary the scale height $\sqrt{\overline{\Phi}}$ and the local radiative equilibrium amplitude $\Delta \Phi_{\rm eq}/\overline{\Phi}$. While the complete ensemble includes a range of radiative timescales and rotation periods, here we chose to highlight the shortest-timescale regime (with radiative timescale $\tau_{\rm rad}=0.1$ days and rotation period $P_{\rm rot}=1$ day) in order to examine the effects of forcing strength.\footnote{We show the geopotential and mean-zonal wind grids for all scale heights and forcing regimes in our Zenodo repository (The DOI for the repository will be provided after peer review)}. Figure \ref{fig:geopot_grid_fig} shows the geopotential contours and Figure \ref{fig:wind_grid_fig} shows the mean zonal wind profiles for each combination of forcing and reference geopotential $\overline{\Phi}$. Note that as we reduce the scale height (and hence $\overline{\Phi}$), the magnitude of forcing decreases, since $\Delta \Phi_{\rm eq}/\overline{\Phi}$ depends on $\overline{\Phi}$.

    In the strong-forcing regime (left column of Figure \ref{fig:geopot_grid_fig}), as we decrease the scale height, the patterns in atmospheric dynamics remain qualitatively similar to the one for $\overline{\Phi}=4\times 10^6$, shown in Figure \ref{high-Phibar-4}. However, the east/west and north/south gradients become more defined as the scale height decreases.
    The mean zonal winds for $\overline{\Phi}=3 \times 10^6$ m$^2$s$^{-2}$ are similar in speed and amplitude to those occurring at higher scale height, ranging from $\approx 10-20$ ms$^{-1}$ for the longest timescales to $\approx 600$ ms$^{-1}$ for the shortest timescales. As we continue to decrease the scale height to $\overline{\Phi}=2 \times 10^6$ m$^2$s$^{-2}$, the qualitative behavior remains the same while the geopotential amplitude and the mean zonal wind speed decrease further, ranging from $\approx 10-20$ ms$^{-1}$ for the longest timescales to $\approx 500$ ms$^{-1}$ for the shortest timescales.  For the lowest scale height $\overline{\Phi}=1 \times 10^6$ m$^2$s$^{-2}$, the wind speeds vary from $\approx 300$ ms$^{-1}$ for the shortest-timescale regime to $\approx 10-20$ ms$^{-1}$ for the longest timescales regime. As the scale height decreases, the hot spot in the short time scales regime (where $\tau_{\rm rad}=0.1$ days, $P_{\rm rot}=1$ day shown in the left column of Figure \ref{fig:geopot_grid_fig}) changes its shape from one matching hot Jupiters (top left panel) to a more elongated in the east-west direction due to rotational effects (bottom left panel). This change is driven by a decrease in Rossby deformation radius, which is proportional to the scale height and inversely proportional to the Coriolis parameter. Overall, decreasing the scale height results in decreased geopotential contrast and more pronounced rotational effects, both due to the relationship between scale height and forcing.

    In the middle column of Figure \ref{fig:geopot_grid_fig}, we show the short-timescale simulations for the medium-forcing regime, where the prescribed local radiative equilibrium day-night contrast is equal to half of the reference geopotential ($\Delta \Phi_{\rm eq}/\overline{\Phi}=0.5$). While the strong forcing regime ($\Delta\Phi_{\rm eq}/\overline{\Phi}=1$) has been widely explored in literature in the context of hot Jupiters \citep[e.g.][]{perez2013atmospheric,showman2011equatorial}, the medium forcing regime has not been investigated as thoroughly \citep[but appears, e.g., in][]{liu2013initialconditions}. This regime is relevant to the modeling of exoplanet atmospheres since many sub-Neptunes orbit closely around cooler stars, such as M-dwarfs, and are synchronous rotators \citep{bean2021nature}. Such sub-Neptunes are also advantageous targets for atmospheric characterization with observational facilities due to the favorable ratio between the planet and star radii \citep{smallstar}.

    The trends in qualitative behavior of these simulations are similar to those for the high contrast (strong forcing) regime. The shortest radiative and rotational timescales result in high equator-to-pole and day-night contrast and highest winds, while the longest-timescale regime exhibits low geopotential contrast and low winds. The amplitude of the resulting geopotential profile appears to scale linearly with the day-night contrast of the local radiative equilibrium. While strong-forcing simulations exhibit a single equatorial hotspot, medium-forcing simulations tend to exhibit two regions of highest geopotential in the mid-latitudes. Despite qualitative similarities, there are distinct differences that arise when comparing to the strong forcing regime.

    Compared to strongly forced regimes at the same scale height, the medium-forcing simulations tend to exhibit similar qualitative trends in response to changing radiative timescale and rotation period. However, there are several quantitative differences. The overall geopotential contrast tends to be smaller. The greatest geopotential contrast is $\approx 6.4 \times 10^5$ m$^2$s$^{-2}$ (for $\tau_{\rm rad}=0.1$ days, $P_{\rm rot}=1$ day). Compared to the strong-forcing regime, the hotspot for the medium-forcing regime extends farther west at mid-latitudes (30$^\circ$-60$^\circ$). The lowest geopotential contrast is $\approx 2 \times 10^2$m$^2$s$^{-2}$ (for $\tau_{\rm rad}=10$ days, $P_{\rm rot}=5$ days). The qualitative wind speed trends for medium-forcing models follow those for strong forcing, but the speeds are significantly lower (with the mean zonal wind speeds ranging from $\approx 500$ ms$^{-1}$ to $\approx 50$ ms$^{-1}$). Overall, the atmospheric dynamics of the medium forcing regime exhibit lower day-night contrasts, weaker equatorial winds speeds, and different location of vortices, compared to the strong forcing regime. However, the transitions between day-to-night flow and jet-dominated flow occur at the same points in the $\tau_{\rm rad}$---$P_{\rm rot}$ parameter space as for the strong forcing regime.

    As scale height decreases, for shortest timescales  (for $\tau_{\rm rad}=0.1$ days, $P_{\rm rot}=1$ day), the hotspot is elongated in the east-west direction at mid-latitudes, and secondary eastward jets emerge at high latitudes in addition to the equatorial jet. The elongation is due to the decrease in the Rossby radius of deformation, which is proportional to $\sqrt{\overline{\Phi}}$. Compared to the same scale height regime at strong forcing (left column of Figure \ref{fig:geopot_grid_fig}), both geopotential contrast and wind speeds are lower, while the jet structure and transitions from day-night to jet-dominated flow are preserved.  In comparison to the strong forcing regime at the same scale height, this regime exhibits lower geopotential contrast ($\approx 50 \%$ decrease) and lower wind speeds (ranging from  $\approx 20$ ms$^{-1}$ for the longest timescales to $\approx 150$ m$\rm{s}^{-1}$ for the shortest timescales).

   In the right column of Figure \ref{fig:geopot_grid_fig}, we present the weak-forcing simulations for the short-timescales regime. For these simulations, the local radiative equilibrium amplitude is equal to 0.1 of the mean geopotential ($\Delta \Phi_{\rm eq}/\overline{\Phi}=0.1$), approaching the linear regime commonly assumed for Earth. Low-contrast regimes have been previously explored in the context of hot Jupiters \citep[e.g.][]{perez2013atmospheric, shell2004abrupt} due to their analytical tractability. Similarly to the medium-forcing regime, this regime models sub-Neptunes orbiting closely around cooler stars. For these weak-forcing simulations, the rotational effects become more pronounced, and the transitions between day-to-night flow and jet-dominated flow occur at shorter radiative timescales compared to medium- and strong-forcing regimes.

    For large scale height ($\overline{\Phi}=4 \times 10^6$ m$^2$s$^{-2}$),  the maximal geopotential contrast is $\approx 1.75 \times 10^5$ m$^2$s$^{-2}$, and mean zonal wind speeds are in the $\approx 30-100$ ms$^{-1}$ range. When $\tau_{\rm rad}=0.1$ days, we observe day-to-night flow, similar to strong and medium forcing regimes, but instead of a single hotspot east of the substellar point we observe two dayside anticyclonic vortices west of the substellar point. The short-timescales regime ($\tau_{\rm rad}=0.1$ days, $P_{\rm rot}=1$ day) shown in the right column of Figure \ref{fig:geopot_grid_fig} exhibits behavior qualitatively similar to the regime where $\tau_{\rm rad}=1$ day, $P_{\rm rot}=1$ day for strong and medium forcing (e.g. middle column of Figure \ref{high-Phibar-4}). This regime is a transition between day-to-night and jet-dominated flow. The similarity occurs because increasing $\tau_{\rm rad}$ has the same effect on Equation \ref{eq:Q} as decreasing $\Delta \Phi_{\rm eq}/\overline{\Phi}$. 
   
     As we decrease the scale height in the weak-forcing regime, we observe similar qualitative behavior, with transitions occurring at the same points in the parameter space. For $\overline{\Phi}=3 \times 10^6$ m$^2$s$^{-2}$, the highest geopotential contrast decreases to $\approx 1.5 \times 10^5$, and the mean zonal wind speeds decrease slightly, resulting in $\approx 25-100$ ms$^{-1}$ range. For $\overline{\Phi}=2 \times 10^6$ m$^2$s$^{-2}$, the maximal geopotential contrast is $\approx 1.1 \times 10^5$ m$^2$s$^{-2}$. The mean zonal winds decrease to $\approx 10$ ms$^{-1}$ in the shortest timescales regime ($\tau_{\rm rad}=0.1$ days and $P_{\rm rot}=1$ day) and  $\approx 80$ ms$^{-1}$ in the longest timescales regime ($\tau_{\rm rad}=10$ days and $P_{\rm rot}=10$ days). Finally, at the lowest scale height in our ensemble, $\overline{\Phi}=10^6$ m$^2$s$^{-2}$, the maximal geopotential contrast for the weak forcing regime is $\approx 7 \times 10^4$ m$^2$s$^{-2}$. The mean zonal winds are in the $\approx 10-80$ ms$^{-1}$ range. 
     
     Overall, compared to strong and medium forcing simulations, the weak forcing regime exhibits lower geopotential contrast and lower wind speeds. The largest geopotential heights are in mid-latitudes, not at the equator, which is distinct from the strong- and medium-forcing regimes. The transitions between day-night flow and jet-dominated flow occur at shorter radiative timescales compared to the strong- and medium-forcing regimes. Note also that for strong and medium forcing, the highest mean zonal wind speeds occur when both radiative timescale and rotation period are short, and the lowest wind speeds occur when both timescales are long. This pattern is reversed in the weak forcing regime. We discuss this transition further in \ref{sec:disc-wind-trends}. Weak forcing induces lower wind speeds, which allow non-advective properties such as rotational effects to dominate, especially at long radiative timescale $\tau_{\rm rad}=10$ days. Here we have noted several patterns that take place in the atmospheres of these planets. In the following section, we discuss general trends and special cases found in our simulations.

\section{Discussion} \label{sec:discussion}
  
     \subsection{Holistic discussion of trends in the regimes}
     \label{sec:holistic-discussion}
     In this section, we discuss the overall trends in the atmospheric patterns in our simulations, focusing on trends in geopotential height distributions. Since atmospheric heating will generally increase the geopotential $\Phi$ and atmospheric cooling will generally reduce $\Phi$, the qualitative changes in geopotential can be thought of as changes in temperature, with higher thickness of the upper layer corresponding to higher temperatures. Using the ideal gas relationship $\Phi \approx R T$, where $R$ is the specific gas constant, we can approximate changes in temperature. For our range of mean molecular weights, the approximate range for values of $R$ is 750---1800 J kg$^{-1}$K$^{-1}$. We can estimate that, for example, a change of $10^5$ m$^2$s$^{-2}$ in the geopotential can correspond to an approximate temperature change in the $ 50-130$ K range, depending on assumed mean molecular weight.
    The geopotential and horizontal winds are driven by the interacting processes of irradiation, rotation, and advection.
     
     The strength of stellar irradiation in our simulation can be analyzed through two parameters: the radiative timescale $\tau_{\rm rad}$ and the day/night temperature contrast of the local radiative equilibrium. The radiative timescale is one of the principal determiners of the resulting atmospheric patterns.
     Simulations at the shortest radiative timescale tend to exhibit a hot spot on the day side and cyclonic vortices on the nightside. The hot spot is shifted eastward from the substellar longitude, similar to the equatorial superrotation on hot Jupiters. High day-night heating contrasts result in stationary, planetary-scale Rossby and Kelvin waves as shown in \citet{showman2011equatorial}.  As the timescale $\tau_{\rm rad}$ increases to 1 day, the hot spot curves westward in the mid-latitudes under the influence of the Coriolis force, breaking up into cyclonic vortices that equilibriate west of the substellar point. As the radiative timescale $\tau_{\rm rad}$ increases further to 10 days, the behavior transitions from a day-to-night flow to a jet-dominated flow. Rossby gyres tend to be present for small and medium values of $\tau_{\rm rad}$ but not for $\tau_{\rm rad}=10$ days. Time variability in our simulations occurs when either the radiative timescale or the rotation period is long.
     
     The other parameter controlling the strength of irradiation is the prescribed equilibrium day/night temperature contrast. The resulting day-night geopotential thickness contrast appears to scale linearly with the fractional day-night equilibrium difference $\Delta \Phi _{\text{eq}}/\overline{\Phi}$. This behavior makes sense as a response to linear Newtonian relaxation. The fractional day-night difference controls the amplitude of the dayside cosine bell of the prescribed local radiative equilibrium. Similarly, as we lower the scale height, lower mean geopotential $\overline{\Phi}$ corresponds to lower resulting day-night thickness contrasts. The equator-to-pole contrast decreases as the radiative timescale or the rotation period decrease.  In general, simulations with larger rotation periods tend to exhibit lower day/night geopotential contrast. Table \ref{table:phibar-4-summary} shows the difference between average dayside and average nightside geopotential. When $\tau_{\rm rad}=1$, for medium and weak forcing, minimum day/night contrast $\Delta \Phi$ occurs for $P_{\rm rot}=5$ days rather than $P_{\rm rot}=10$ days. This behavior occurs because the contributions from the hotspot or cyclonic vortices are split more evenly between the dayside and the nightside in these regimes. For shorter and longer $\tau_{\rm rad}$, the hotspot is predominantly on the dayside. However, $\Delta \Phi$ provides a useful heuristic for order-or-magnitude comparisons of day/night contrasts in our ensemble. 
     
     The response in our simulations to changes in radiative forcing is consistent with the argument given in \citet{showman2012doppler}. When radiative forcing is weak (e.g. in the weak forcing regime or due to longer $\tau_{\rm rad}$), the heating of air parcels at they cross from the day side to the night side will be too small to induce significant day-night contrast, and the dominant driver of the flow will be the meridional (latitudinal) gradient in the zonal-mean radiative heating. As expected, we generally observe zonal symmetry with primary temperature variations occurring between the poles and the equator.

     The horizontal wind speeds in our simulations are controlled by radiative forcing. In the strong and medium forcing regimes, the largest mean zonal wind speeds tend to occur when both the radiative time scale and the rotation period are short. In this case, the mean zonal wind speeds form a global-scale eastward equatorial jet that moves with the speed of a few hundred ms$^{-1}$. For longer timescales $\tau_{\rm rad}$ and $P_{\rm rot}$, the mean zonal wind speeds are $\overline{U}\approx 100$ ms$^{-1}$. In the weak forcing regime, the mean zonal wind speeds are generally lower and can drop to  $\overline{U}\approx 10$ ms$^{-1}$. In the weak forcing regime, the mean zonal winds tend to be lower when the planetary rotation period is short. We discuss this behavior in more detail in Section \ref{sec:disc-wind-trends}.
     
\begin{figure}
  \centering
  \includegraphics[width=\linewidth]{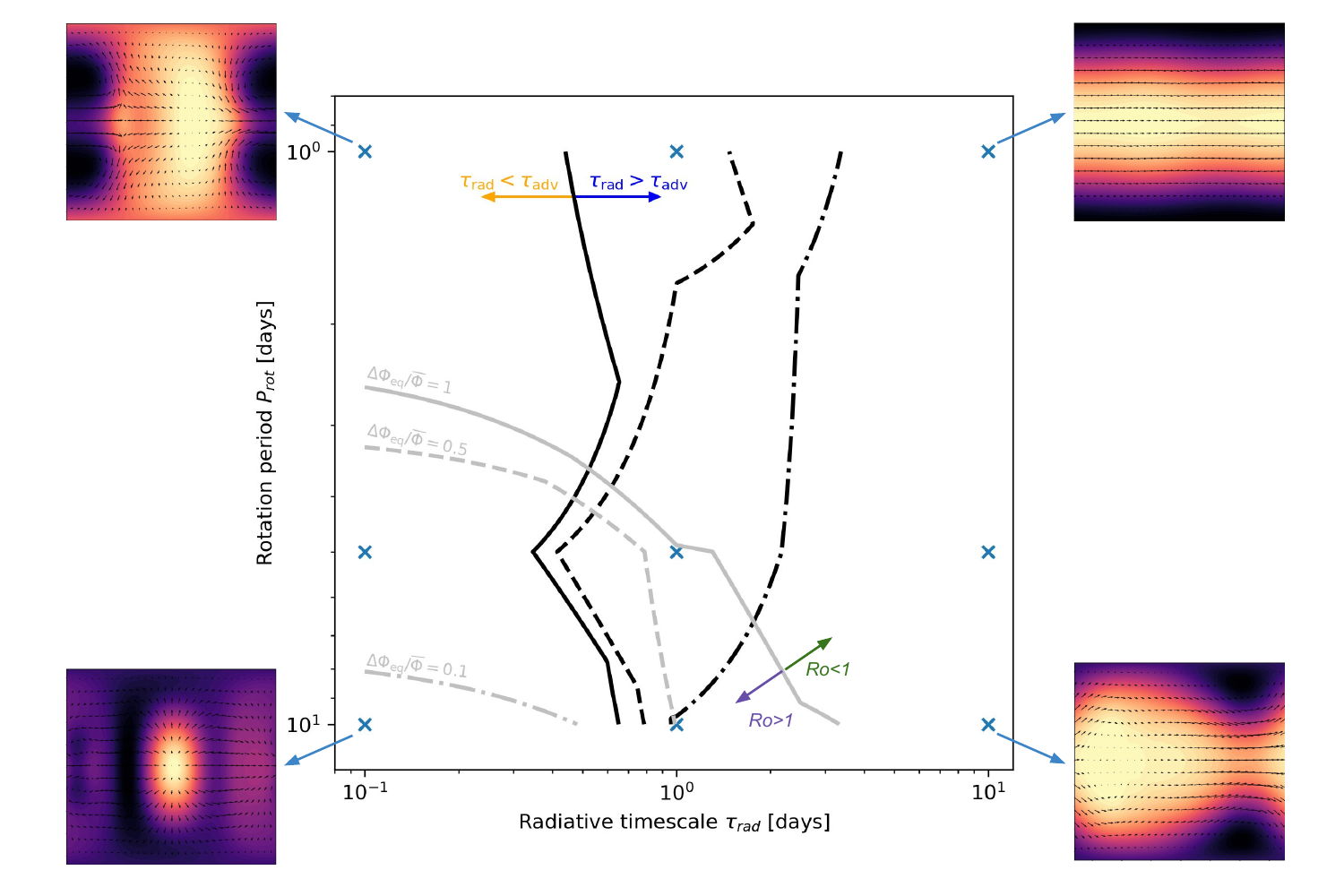}
  \caption{The estimated contours where $\tau_{\rm adv}=\tau_{\rm rad}$ (black) and the estimated contours where $Ro=1$ (gray) for different day/night contrasts of the local radiative equilibrium for the strong (solid line), medium (dashed line), and weak (dot-dashed line) forcing regimes. The blue crosses mark the simulated regimes. In order to compute the contours, the values for $\tau_{\rm rad}$, $\tau_{\rm adv}$, and $Ro$ were averaged across the four scale heights in our ensemble. Example patterns are presented in the four corners of the parameter space. The color scale is normalized individually for each panel in order to highlight atmospheric patterns.}
  \label{tau_adv_contours}
\end{figure}

     We now turn to consider the balance of advective and radiative forces using scale analysis. Here we compare the radiative timescale $\tau_{\text{rad}}$ and the advective timescale $\tau_{\rm adv}=L/U$, where $L$ is a characteristic horizontal length scale. We use $L\sim a$ for a global-scale flow.  For regimes where $\tau_{\text{rad}}=0.1$ days and $\tau_{\text{rad}}=1$ day, $\tau_{\text{rad}}<\tau_{\text{adv}}$, and so the day-night differences tend to be higher. However, when $\tau_{\text{rad}}=10$, we tend to observe $\tau_{\text{rad}}>\tau_{\text{adv}}$, where the advective behavior dominates exhibiting planetary-scale westward Rossby waves. These waves are especially dominant for regimes where $P_{\text{rot}}=\tau_{\text{rad}}=10$ days.
    For the intermediate scale $\tau_{\text{rad}}=1$ day, the strong forcing regime tends to have $\tau_{\text{rad}}>\tau_{\text{adv}}$, while the weak forcing regime tends to have $\tau_{\text{rad}}<\tau_{\text{adv}}$, and the medium forcing regime is transitional. This behavior is summarized in Figure \ref{tau_adv_contours}. In black, the figure shows estimated contours where  $\tau_{\text{rad}}=\tau_{\text{adv}}$ computed via interpolation of the values for each of the strong-forcing (solid line), medium-forcing (dashed line), and weak-forcing (dash-dotted line) regimes. The values were averaged over all reference values of $\overline{\Phi}$. When the advective timescale is longer than the radiative timescale, the simulations exhibit high day-night contrasts. When the advective timescale is shorter than the radiative timescale, the simulations exhibit little longitudinal variation.

    The importance of rotation varies across our simulations. The strong and medium forcing regimes tend to be driven primarily by insolation, while in the weak forcing regime, rotation has a more pronounced effect. {We quantify the dominance of rotation using the Rossby number $Ro=U/2\Omega L$, which gives the ratio of advective to Coriolis forces away from the equator in the horizontal momentum equation.} $U$ is the characteristic windspeed and $L$ is the characteristic horizontal length scale.
    Rapidly rotating planets with $P_{\rm rot}=1$ day and planets with $\tau_{\rm rad} =10$ days exhibit $Ro \ll 1$ regardless of irradiation. The highest Rossby numbers ($\sim 2$) are exhibited by slowly rotating planets with short radiative timescales. When rotation period is 5 or 10 days and the radiative timescale $\tau_{\rm rad}=1$ day, irradiation plays more of a role in influencing the Rossby number. Strongly forced planets tend to exhibit $\rm{Ro} \sim 1$, weakly forced planets exhibit $Ro \ll 1$, and planets with medium forcing exhibit a transition between these two regimes. Figure \ref{tau_adv_contours} shows the estimated contours where  $Ro = 1$ computed via interpolation of the values for each of the strong-forcing (solid line), medium-forcing (dashed line), and weak-forcing (dash-dotted line) regimes in gray. The values were averaged over all reference values of $\overline{\Phi}$. As expected, the effects of rotation are more dominant in the atmospheres of rapidly rotating planets and planets with lower insolation, resulting in a chevron-shaped hotspot when combined with short radiative timescale $\tau_{\rm rad}$, and in jet-dominated flow when combined with long $\tau_{\rm rad}$. When the radiative timescale is short and the rotation period is long, the steady-state geopotential resembles the local radiative equilibrium. Overall, the strong and medium forcing regimes are similar to each other, while the weak forcing regime exhibits lower Rossby numbers. Changes in scale height do not affect the Rossby number significantly. Rossby numbers for a representative scale height corresponding to $\overline{\Phi}=4\times 10^6$ m$^2$s$^{-2}$ are represented in Table \ref{table:phibar-4-summary}. 
    
    Due to low insolation, the dynamics of the weak forcing regime tend to be qualitatively different from the dynamics of the strong and medium forcing regimes at the same timescales. Low stellar irradiation results in lower wind speeds and lower contrasts, as well as the formation of jet-dominated flow. {Instead of eastward superrotation, the weak forcing regimes tend to exhibit westward shifts of maximal geopotential driven by Rossby waves.} The interaction of low insolation and rotation effects results in lower longitudinal variation than the strong- and medium-forcing regimes.
    
    Rossby radius of deformation is greater than the planetary radius for $P_{\text{rot}}=5$ days and $P_{\text{rot}}=10$ days.{This estimate predicts the formation of global    scale flows in the corresponding regimes. This prediction is confirmed by our simulations, which exhibit global scale flows in both geopotential and horizontal wind patterns.} For $P_{\text{rot}}=1$ day, the Rossby radius of deformation ranges from $6.8 \times 10^6$ for $\overline{\Phi}=10^6$ to $1.3 \times 10^7$ for $\overline{\Phi}=4 \times 10^6$. This corresponds to lower widths of the equatorial jets, which extend up to $20^{\circ}-45^{\circ}$ latitude. Compared with our simulations, this estimate tends to hold for $\tau_{\rm rad}=0.1$ days, $P_{\rm rot}=1$ day, but as the radiative timescale increases, the equatorial jet becomes suppressed. We discuss this behavior in further detail in Section \ref{sec:disc-wind-trends}.

    \subsection{Trends in the zonal wind flow}
    \label{sec:disc-wind-trends}
    
    In our simulations, we have identified several important trends in the zonal wind flow, which we discuss here. In the strong forcing, high scale height regime, the shape of the mean zonal winds depends strongly on the interaction of the rotation period and radiative timescale. For example, for high scale height and strong forcing, in the top row of Figure \ref{high-Phibar-4-wind}, the shortest timescales regime ($P_{\rm rot}=1$ day, $\tau_{\rm rad}=0.1$ days) exhibits a strong equatorial jet. However, as the radiative timescale increases to $\tau_{\rm rad}=1$ day and $\tau_{\rm rad}=10$ days, the equatorial jet becomes suppressed and replaced by two symmetrical jets with peaks at $\approx 40^{\circ}$ latitude. As $\tau_{\rm rad}$ increases, the forcing weakens. This in turn weakens the equatorial jet, which typically emerges as a response to longitudinal variation in radiative forcing \citep[e.g.][]{showman2011equatorial}. In these conditions, strong Coriolis effect results in the emergence of symmetric tropical waves, similar to what is observed in idealized models of Earth troposphere \citep[e.g.][]{kraucunas2007tropical}. 
    
    The transition from a single equatorial jet to symmetric tropical jets appears only in the presence of a short rotation period. Note that in the middle row of Figure \ref{high-Phibar-4-wind} (where the rotation period $P_{\rm rot}=5$ days), the symmetric tropical jet pattern is weaker. In the bottom row (at rotation period $P_{\rm rot}=10$ days), the equatorial jet weakens but does not disappear as $\tau_{\rm rad}$ increases. Our simulations suggest that in the sub-Neptune regime, the rotation period must be sufficiently short and the radiative timescale must be sufficiently long in order to see symmetric tropical jets.
    
    The emergence of symmetric tropical jets persists across scale height variations in the high-forcing and medium-forcing regimes. As scale height decreases, the overall mean zonal wind speed decreases, and the transitions in qualitative behavior occur at longer values of $\tau_{\rm rad}$. {For example, the simulation for strong forcing at low scale height exhibits an equatorial jet when $\tau_{\rm rad}=1$ day, $P_{\rm rot}=5$ days, while the strong forcing simulations at high scale height exhibit tropical jets at these timescales.} The medium-forcing regime exhibits similar qualitative behavior as the strong-forcing at low scale height at all the simulated scale heights, except for $\tau_{\rm rad}=10$ days and $P_{\rm rot}=5$ or $10$ days, where we observe a pattern which is neither a strong equatorial jet nor symmetric tropical jets. Here a single westward jet has a flat peak that extends across the equator, exhibiting an intermediate, transitional regime. As in the geopotential patterns, the wind patterns for high- and medium-forcing regimes exhibit more similarities to each other than to the winds in the weak-forcing regime.
    
    The shape of the jet pattern at shortest timescales ($\tau_{\rm rad}=0.1$ days, $P_{\rm rot}=1$ day) changes in response to changes in forcing. In the strong-forcing regime, when the scale height is large (at $\overline{\Phi}=4\times10^6$ m$^2$/s$^2$), the simulation exhibits westward jets north and south of the eastward equatorial jet {(shown in the top left panel of Figure \ref{fig:wind_grid_fig})}. The peaks of the westward jets correspond to the minima of mean zonal winds $\overline{U}$, which occur at $\approx 50^{\circ}$ latitude. However, as scale height decreases, the minima move toward the equator. This behavior is a response to the narrowing of the equatorial jet, since decreasing the scale height also decreases the Rossby radius of deformation ($\lambda_R=\sqrt{gH}/f=\sqrt{\overline{\Phi}}/f$). For $\overline{\Phi}=10^6$ m$^2$/s$^2$, the minima occur at $\approx 35^{\circ}$ latitude  {(bottom left panel of Figure \ref{fig:wind_grid_fig}).} {In this regime, secondary eastward jets emerge polarward of the westward jets. The eastward jets are similar in speed to the westward jets ($\approx 35$ ms$^{-1}$).} In the medium-forcing regime ( {middle column of Figure \ref{fig:wind_grid_fig}}), the secondary jets are present at both $\overline{\Phi}=10^6$ ($\approx 35$ ms$^{-1}$) and $\overline{\Phi}=2\times 10^6$ (with speeds of $\approx 20$ ms$^{-1}$) and are much slower than the eastward equatorial jet. In contrast, in the weak-forcing regime ( {right column of Figure \ref{fig:wind_grid_fig}}), at shortest timescales, the secondary jets are of the same order of magnitude as the eastward equatorial jet.   The combination of low rotation period and weak forcing produces a zonal wind pattern similar to that exhibited by the models of hot Jupiter WASP-43b in presence of high drag  \citep{kataria2015atmospheric}.

     \subsection{Spin-up behavior for long timescales}
     \label{sec:wandering-hot-spot}
\begin{figure}
  \centering
  \includegraphics[width=0.5\linewidth]{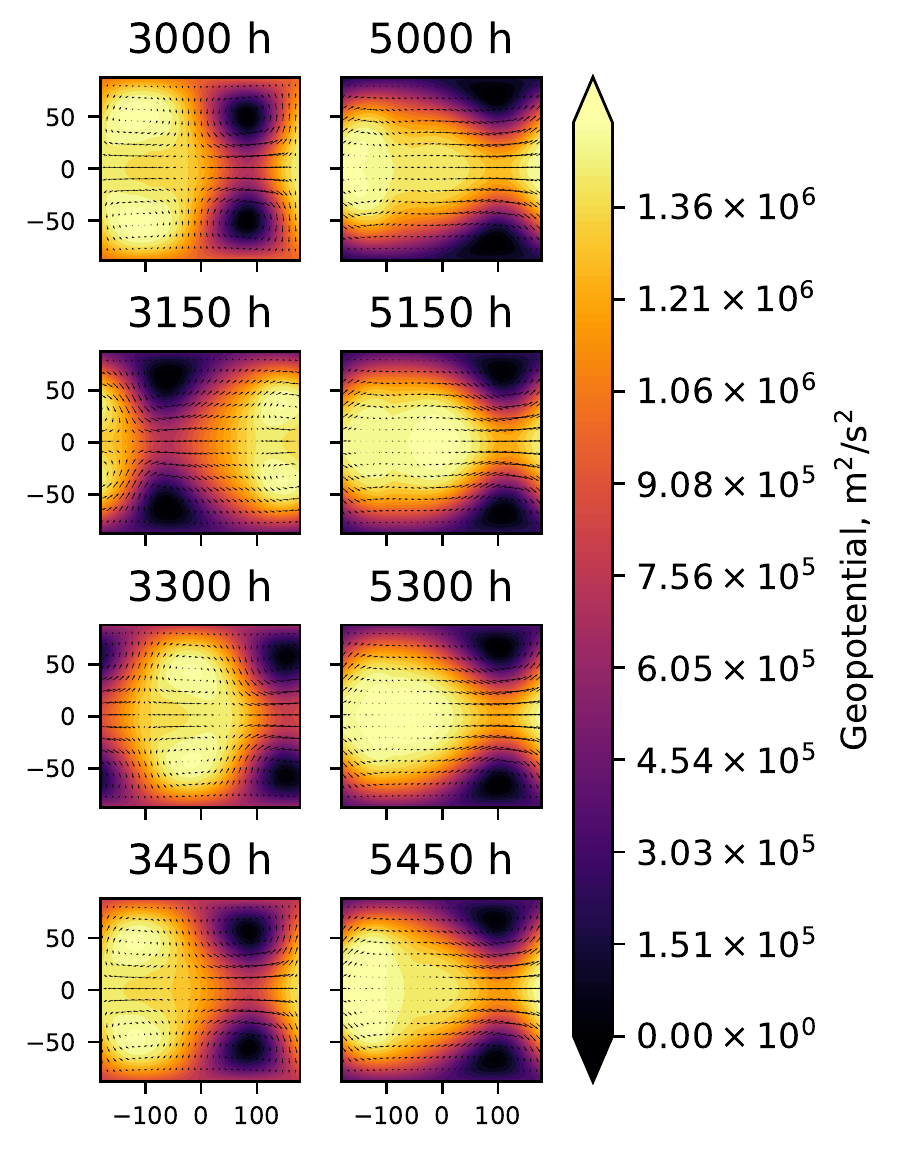}
  \caption{Snapshots of spin-up behavior for strong-forcing regime with $\overline{\Phi}=4\times10^6$ at rotation period $P_{\rm rot}=10$ days and radiative timescale $\tau_{\rm rad}=10$ days. The hotspot is advected fully around the planet during spin-up (left column shows cyclic behavior with a period of $\approx 450$ hours). The right column shows the equilibrating behavior. Time variability diminishes.} 
  \label{spinup-advection}
\end{figure}

     When both $\tau_{\rm rad}$ and $P_{\rm rot}$ are long, the model exhibits long spin-up behavior driven by the westward Rossby wave. During spin-up, the hot spot can be advected around the planet several times before the oscillations subside. This behavior is illustrated in Figure \ref{spinup-advection} for $\overline{\Phi}=4\times 10^6$ in a strong forcing regime, but occurs for long radiative timescales for a variety of scale heights in the strong and medium forcing regimes. The left column of Figure \ref{spinup-advection} captures the Rossby wave advecting the hotspot westward all the way around the planet with a period of $\sim 450$ hours. However, this oscillation is transient, and eventually the hot spot is west of the substellar point and is no longer advecting around the planet. This behavior highlights the need for longer simulations, especially at long timescales and more temperate forcings, where time variability is more likely. 
   
   \subsection{North-south hemisphere asymmetry and apparent pitchfork bifurcation}
   \label{sec:pitchfork}
   
     \begin{figure}
  \centering
  \includegraphics[width=0.5\linewidth]{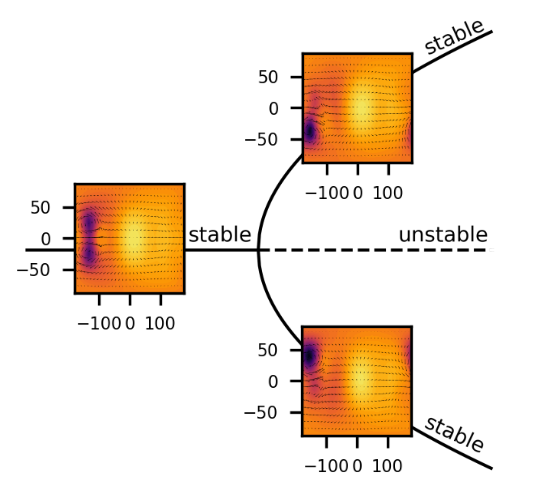}
  \caption{Supercritical pitchfork bifurcation diagram illustrating a potential mechanism for the emergence of asymmetric gyres. Prior to the bifurcation, the symmetric equilibrium is stable. At the bifurcation point, the previously stable symmetrical equilibrium becomes unstable, and two new equilibrium branches appear, with one cyclonic gyre in either north or south hemisphere. } 
  \label{pitchfork-diagram}
\end{figure}

   Several simulations in our ensemble exhibit north-south hemisphere asymmetry, which typically manifests as differences between Rossby gyres. In some regimes, the asymmetric Rossby gyres are stationary, while other regimes exhibit oscillations, or migrating gyres.  {Both of these phenomena tend to occur when the radiative timescale is short and the rotation period is long, and in the transitional regimes between the short-timescale day-to-night flow and the long-timescale jet-dominated flow.}
   
   {The Rossby gyres are stationary for $\tau_{\rm rad}=0.1$ days and $P_{\rm rot}=10$ days for medium forcing at $\overline{\Phi}=3 \times 10^6$ and $4 \times 10^6$ m$^2$s$^{-2}$, and for strong forcing at $\overline{\Phi}=4 \times 10^6$ m$^2$s$^{-2}$.} During spinup, these regimes first approach a symmetric pattern, but at $\sim 2000$ simulated hours, the asymmetry emerges. The transition to the asymmetrical pattern begins near the convergence zone, directly east of the gyres, where the eastward equatorial jet between the gyres meets the westward day-to-night flow. In the convergence zone, a minimal perturbation to the eastward jet is sufficient to begin the transfer of momentum across the equator and destabilize the symmetry.
   
   This behavior is consistent with a supercritical pitchfork bifurcation, shown schematically in Figure \ref{pitchfork-diagram}. When the bifurcation occurs, the symmetric equilibrium (with symmetric nightside gyres) becomes unstable, and two stable asymmetric equilibria (one with a northern, one with a southern gyre) emerge. We have validated the stability of both asymmetric equilibria by reflecting the profile around the equator and using it as the initial condition for another simulation. The bifurcation point appears time-dependent. \citet{jiang1995multiple} describe a qualitatively similar asymmetrical double-gyre behavior in a shallow-water model for an Earth-like context, with wind stress as the bifurcation parameter. However, the planetary parameters are sufficiently different in our study that further mathematical investigation is needed to identify a bifurcation parameter and map the full phase space of these equilibria. 
   
    {Several regimes in our ensemble exhibit persistent oscillations with migrating gyres. These oscillations are present for the strong forcing regime when $P_{\rm rot}=5$ or 10 days, and $\tau_{\rm rad}=1$ day at most sampled scale heights, and for the weak forcing regime when when $P_{\rm rot}=10$ days, and $\tau_{\rm rad}=0.1$ days at all but the smallest scale height. Through Lomb-Scargle analysis, we established that the period of these oscillations is $\approx50$ days for the strong forcing regime, and $\approx30$ days for the weak forcing regime, equal to several multiples of the orbital period. Note that the ratio $\Delta \Phi_{\rm eq}/\tau_{\rm rad}$ is the same for strong forcing when $\tau_{\rm rad}=1$ day and for weak forcing when $\tau_{\rm rad}=0.1$ days.} While the emergence of these oscillations is distinct from the apparent pitchfork bifurcation, the fact that many of the resulting oscillations exhibit hemispheric asymmetry and that they tend to occur at similar combinations of radiative timescale and rotation period suggests that these phenomena are connected. The advection in these regimes appears to oscillate through the two stable asymmetric patterns shown in Figure \ref{pitchfork-diagram}.

    {This behavior is distinct from the typical behavior of 2D exoplanet simulations. Hot Jupiters tend to exhibit a Matsuno-Gill \citep{matsuno1966quasi,gill1980some} type stationary-wave balance between Rossby and Kelvin waves (which can be seen, for example, in the left and center panels of Figure \ref{PBS-recreation}). \citet{showman2009} find a coherent oscillation with a period of 43 days in their simulation the hot Jupiter HD 189733b using SPARC/MITgcm, suggesting a global sloshing mode with a period similar to that of the strong-forcing 50 day oscillations in our study. Further, both variability and north-south asymmetry have been documented in the both 2D and 3D GCM studies of exoplanets. We expand on connections to previous studies in Section \ref{sec:comparisons-literature}.} In both static and oscillatory asymmetrical regimes, the radiative timescale is short and the rotation period is long. Compared to the shorter rotation period of hot Jupiters, the Coriolis force is weaker in this region of our parameter space. The weaker rotational effects appear to allow easier transfer of momentum across the equator. Further investigation of the parameter space in conjunction with computational bifurcation theory tools could illuminate this phenomenon further but falls outside the scope of this work. The simulation snapshots and zonal wind flows presented here are representative of large, global-scale flow patterns.
   
   \subsection{Equatorial and tropical band oscillation}
   \label{sec:tropical-oscillation}
   
\begin{figure}
  \centering
  \includegraphics[width=0.5\linewidth]{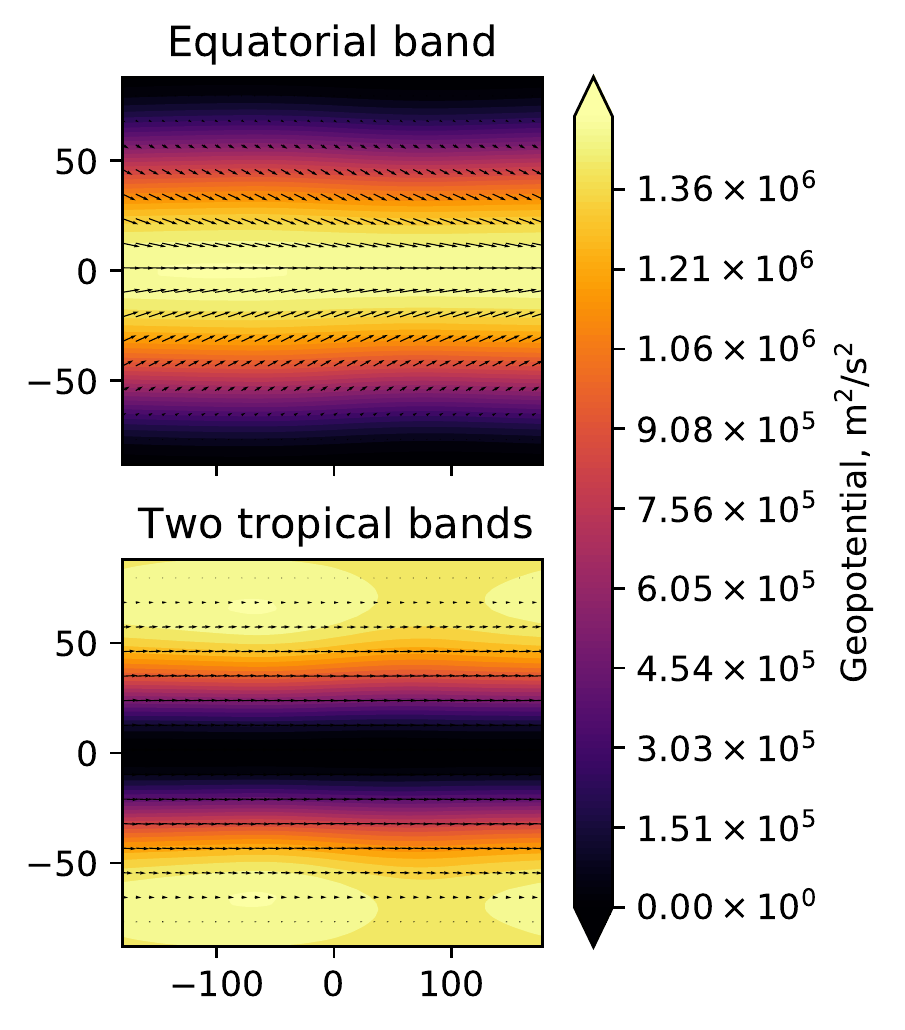}
  \caption{Snapshots of oscillatory behavior for a regime corresponding to a weakly forced planet with a scale height corresponding to $\overline{\Phi}=4\times10^6$ at rotation period $P_{\rm rot}=1$ days and radiative timescale $\tau_{\rm rad}=10$ days. Over the course of the oscillation, a hot equatorial band (top panel) alternates with two hot tropical bands (bottom panel) with a period of $\approx 55$ hours.}
  \label{qbo}
    \end{figure}

   For simulations with a short rotation period ($P_{\rm rot}=1$ day) and long radiative timescale ($\tau_{\rm rad}=10$ days), we observe a persistent oscillation in the strong, medium, and weak forcing regimes with the largest scale height corresponding to $\overline{\Phi}=4 \times 10^4$ m$^2$s$^{-2}$.
   Figure \ref{qbo} shows snapshots of this phenomenon in the low-forcing regime: a hot equatorial band (top panel) alternates with two hot tropical bands (bottom panel). Note that Figure \ref{qbo} shows snapshots corresponding to  the average geopotential profile shown in Figure \ref{high-Phibar-4}.  {The period of these oscillations is $\approx 40$ hours for strong forcing, $\approx 46$ hours for medium forcing, and $\approx 55$ hours for weak forcing.}

     {The oscillation has a structure similar to a gravity-wave like,
    internal, zonally-symmetric Hough mode \citep{longuet1968eigenfunctions}  with
    alternating periods of poleward advection of mass away from the
    equatorial band, followed by equatorward advection of mass. The Coriolis
    force acting on this zonally symmetric meridional flow gives rise to
    oscillations in the zonal mean winds as well. However, the obtained
    period is much longer than the period of the linear mode (about 6.4
    hours). The structure of the quasistationary Kelvin and Rossby waves
    change substantially over the course of the oscillation in the zonal
    mean jet, leading us to speculate that two-way interactions between the
    waves and the mean flow are responsible for destabilizing the steady
    state solution. Further work is needed to verify this speculation.}
   
   The behavior of a single hot equatorial band separating into two hot tropical bands in a persistent oscillation is unique among our simulations and only exists at the largest scale height. When $P_{\rm rot}=1$ days and $\tau_{\rm rad}=10$ days, simulations in all forcing regimes at smaller scale heights show similar qualitative behavior during spinup, but retain the hot equatorial band throughout and are eventually dampened. A large scale height appears necessary for these oscillations to persist.

    \subsection{ {Comparisons with Three-Dimensional Model Studies}}
    \label{sec:comparisons-literature}
     {In this study we explored sub-Neptunes, which bridge the population of hot Jupiters and terrestrial exoplanets. Some of the same phenomena that we see in our simulations have also been observed in 3D GCM simulations of sub-Neptunes as well as hotter gas giant and cooler terrestrial planets. In this section we discuss the connections between our study and existing literature regarding longitudinal variation, equator-to-pole contrast, variability, and the emergence of north-south asymmetry.}

     {Many of the same phenomena and trends that we see evolve in our shallow-water simulations of sub-Neptune atmospheres have appeared in previous studies of sub-Neptunes with 3D GCMs. Many such studies explored the atmospheric physics of one of the earliest discovered warm sub-Neptunes GJ~1214b \citep[e.g.][]{Menou2012, kataria2011, charnay2015, drummond2018}. Although the above models were tuned to a specific planet under specific insolation conditions, both \citet{charnay2015} and \citet{kataria2011} note that Neptunes with higher molecular weight atmospheres tend to exhibit narrower equatorial jets.  \citet{zhang2017effects} confirm this trend in a parameter sweep investigating the effects of molecular weight and molar heat capacity. This behavior is consistent with our simulations, where high molecular weight corresponds to lower scale heights and lower values for the Rossby radius of deformation.}
    
     { \citet{innes2022} also aim at mapping the parameter space and diagnosing the underlying dynamical mechanisms taking place in sub-Neptune atmospheres. \citet{innes2022} explore the dry dynamics of temperate sub-Neptune planets with rotation periods of 6 and 33 days and insolation ranging from half to twice the insolation of Earth. Consistent with our results, their simulations exhibit higher equator-to-pole gradient and lower longitudinal variation for shorter rotation periods. The long (33-day) rotation period simulations also exhibit a cold spot near (but offset from) the poles, which we observe in the weak forcing regime for $\tau_{\rm rad}=1$ day. This correspondence suggests that this pattern is due to a balance between rotational effects and insolation.  }

     {Several atmospheric circulation patterns that we have identified in our ensemble appear in 3D GCM simulations of terrestrial planets. First, in our study, we observe the emergence of symmetric tropical jets when the radiative constant is long and the rotation period is short, which is the closest to terrestrial forcing among our simulations. The emergence of symmetric tropical jets has been noted in \citet{kraucunas2007tropical} using a shallow-water model of an idealized Earth-like planet, as well as in \citet{sergeev2022bistability} in the context of bistability of the TRAPPIST-1e system modeled with a 3D GCM. Second, in our study, we observe north-south asymmetry which tends to emerge  for short radiative timescale and medium rotation period, as a transitional regime between a chevron-shape characteristic for a shorter rotation period and a day-to-night flow typical for a longer rotation period. \citet{noda2017circulation} find similar long-timescale north-south asymmetry as a transitional regime between eastward equatorial jet pattern and mid-latitude westerly jet pattern in the simulated dynamics of a synchronously rotating terrestrial aquaplanet.  The north-south asymmetry regime in our study exhibits bistability consistent with a pitchfork bifurcation behavior, in contrast to studies like \citet[e.g.]{edson2011} where the emergence of multiple equilibria is associated with hysteresis. Finally, migrating vortices similar to the behavior in our ensemble have likewise been documented in the context of 3D simulations of tidally synchronized exoplanets in, \citet[e.g.]{skinner2022, cohen2022}. In moist simulations of TRAPPIST-1e, \citet{cohen2022} find migrating Rossby gyres with an oscillation period of $\approx20$ days, which is approaching the $20$- to $30$-day period exhibited by our weak-forcing simulations. The migrating vortices we observe are cyclonic, and consistently with \citet{skinner2022}, they dissipate and re-form. While \citet{skinner2022} describe quasi-periodic behavior, the oscillations in our ensemble are fully periodic after spin-up.}

\section{Conclusions}
\label{sec:conclusion}

   In this study we have explored radiative forcing and rotational regimes relevant to sub-Neptune atmospheres that gave rise to new patterns of interest in the global-scale flow of these atmospheres. In our model we have varied the radiative forcing as a function of equilibrium geopotential contrast as well as a radiative timescale $\tau_{\rm rad}$. We have additionally varied the planetary rotation period $P_{\rm rot}$. We found that decreasing the planetary radius from a hot Jupiter to a sub-Neptune leads to lower geopotential contrast, even at equivalent radiative forcing. This phenomenon can partly be explained by the following mechanism. Reduced planetary radius leads to shorter advective timescales: a parcel of gas moving at the same zonal speed will move around the planet and pass through the dayside more frequently on a smaller planet rather than a larger one. Shorter advective timescales result in more uniform geopotential.

    We have identified the regions in the parameter space where qualitative transitions in global flow patterns take place.
    {The strong and medium forcing regimes tend to exhibit qualitatively similar behavior: day-night circulation for short radiative timescales and jet-dominated flow with little longitudinal variation for long radiative timescales. }
    In contrast, in the weak forcing regime, due to weaker insolation, the resulting patterns are more driven by the Coriolis force and are more likely to exhibit a westward shift of the hot spot as well as a jet-dominated structure.
    The transition from day-to-night flow to jet-dominated flow is tied predominantly to the radiative timescale $\tau_{\rm rad}$, irrespective of rotation period $P_{\rm rot}$. Short radiative timescales lead to a day-to-night flow, while long radiative timescales lead to jet-dominated flow. The medium value of $\tau_{\rm rad}=1$ day is transitional. For short rotation periods, the intermediate value $\tau_{\rm rad}=1$ day leads to jet-dominated flow, but as rotation period increases, these regimes tend to exhibit more longitudinal variation. Planets with a short rotation period have higher equator-to-pole contrasts than planets with a long rotation period, primarily due to lower Rossby radius of deformation. The day-night geopotential contrast tends to scale linearly with the amplitude of the prescribed local radiative equilibrium. The contrast is highest when both the radiative timescale and the rotation period are short, and is lowest when both are long.
     
    We have further identified the qualitative trends in the mean zonal winds. The transition from a single equatorial jet to symmetric tropical jets occurs for low rotation period as the radiative timescale increases. The emergence of secondary high-latitude jets in addition to the equatorial jet occurs in the shortest timescales regime as scale height decreases. Mean zonal wind speeds range from $\sim 600$ ms$^{-1}$ for the strong-forcing regime to $\overline{U}<100$ ms$^{-1}$ for the weak-forcing regime. In the weak-forcing regime, when the rotation period is short, the mean zonal wind speeds can be as low as $\sim 10$ ms$^{-1}$.  {Using estimates for the wind speeds from our model, we compute the Rossby number, which indicates that we should expect global-scale phenomena. This behavior is consistent with our simulations. }
     
    Temporal variability appears when either the radiative timescale $\tau_{\rm rad}$ or the rotation period $P_{\rm rot}$ is long. When both timescales are short, we see almost no variation. With an eye toward observations, $\tau_{\rm rad}$ might be difficult to determine a priori. However, planets at shorter rotation periods are more likely to have short radiative timescales and thus to exhibit less temporal variability, and can therefore be more favorable candidates for observations over many epochs using current facilities.
     
   The code employs the following software packages for scientific computing: \texttt{SciPy} \citep{Virtanen2020}, \texttt{NumPy} \citep{Harris2020}, and \texttt{Matplotlib} \citep{hunter2007}.

This research was carried out (in part) at the Jet Propulsion Laboratory, California Institute of Technology, under a contract with the National Aeronautics and Space Administration and funded through JPL’s Strategic University Research Partnerships (SURP) program. Support for this work was provided out in part by NASA through a grant from the Space Telescope Science Institute, which is operated by the Association of Universities Research in Astronomy, Inc. under NASA contract NAS5-26555. Support for program HST-GO-16194 was provided through a grant from the STScI under NASA contract NAS5-26555.

\bibliography{main}{}
\bibliographystyle{aasjournal}

\end{document}